\documentclass[prd,amsmath,notitlepage,twocolumn]{revtex4-2}
\usepackage{graphicx}
\usepackage{float}
\usepackage{epstopdf,cancel}
\usepackage{epsf,latexsym,bbm,euscript}
\usepackage{amssymb,amsmath}
\usepackage{mathtools} 
\usepackage{times,graphics}
\usepackage{soul,xcolor}
\usepackage{mathtools}

\usepackage{url,hyperref}
\hypersetup{colorlinks,linkcolor={blue!55!black},citecolor={red!50!black},urlcolor={blue!45!black},breaklinks=true}

\usepackage{scalerel}
\usepackage{tikz}
\usetikzlibrary{svg.path}
\definecolor{orcidlogocol}{HTML}{A6CE39}
\tikzset{
	orcidlogo/.pic={
		\fill[orcidlogocol] svg{M256,128c0,70.7-57.3,128-128,128C57.3,256,0,198.7,0,128C0,57.3,57.3,0,128,0C198.7,0,256,57.3,256,128z};
		\fill[white] svg{M86.3,186.2H70.9V79.1h15.4v48.4V186.2z}
		svg{M108.9,79.1h41.6c39.6,0,57,28.3,57,53.6c0,27.5-21.5,53.6-56.8,53.6h-41.8V79.1z M124.3,172.4h24.5c34.9,0,42.9-26.5,42.9-39.7c0-21.5-13.7-39.7-43.7-39.7h-23.7V172.4z}
		svg{M88.7,56.8c0,5.5-4.5,10.1-10.1,10.1c-5.6,0-10.1-4.6-10.1-10.1c0-5.6,4.5-10.1,10.1-10.1C84.2,46.7,88.7,51.3,88.7,56.8z};
	}
}

\newcommand\orcidlink[1]{\href{https://orcid.org/#1}{\mbox{\scalerel*{
				\begin{tikzpicture}[yscale=-1,transform shape]
					\pic{orcidlogo};
				\end{tikzpicture}}{X}}}}

\def\6{{\langle}}
\def\9{{\rangle}}
\newcommand{\defeq}{\vcentcolon=}
\newcommand{\eqdef}{=\vcentcolon}

\newcommand{\be}{\begin{equation}}
\newcommand{\ee}{\end{equation}}
\newcommand{\ba}{\begin{eqnarray}}
\newcommand{\ea}{\end{eqnarray}}

\def\eD{\EuScript{D}}

\def\half{{\tfrac{1}{2}}}

\def\pad{{\partial}}

\def\ha{{\hat{a}}}

\usepackage[caption=false]{subfig}
\newcommand{\subfigimg}[3][,]{%
  \setbox1=\hbox{\includegraphics[#1]{#3}}
  \leavevmode\rlap{\usebox1}
  \rlap{\hspace*{-10pt}\raisebox{.5\baselineskip}{\small{#2}}}
  \phantom{\usebox1}
}

\def\sg{\textsl{g}}

\def\cO{\mathcal{O}}

\def\rA{\mathrm{A}}

\def\mfr{\mathfrak{r}}

\begin{document}

\title{Horizon-bound  objects: Kerr--Vaidya solutions }

\author {Pravin K. Dahal} 
\email{pravin-kumar.dahal@hdr.mq.edu.au}

\author{Swayamsiddha Maharana}
\email{swayamsiddha.maharana@hdr.mq.edu.au}

\author{Fil Simovic}
	\email{fil.simovic@mq.edu.au}

\author{Daniel R.\ Terno}
\email{daniel.terno@mq.edu.au}

\affiliation{School of Mathematical and Physical Sciences, Macquarie University, NSW 2109, Australia}

\begin{abstract}

 {Kerr--Vaidya metrics are the simplest dynamical axially-symmetric solutions, all of which violate the null energy condition and thus are consistent with the formation of a trapped region in finite time according to distant observers. We examine different classes of Kerr--Vaidya metrics, and find two which possess spherically-symmetric counterparts that are compatible with the finite formation time of a trapped region. These solutions describe evaporating black holes and expanding white holes. We demonstrate a consistent description of accreting black holes based on the ingoing Kerr--Vaidya metric with increasing mass, and show that the model can be extended to cases where the angular momentum to mass ratio varies.   For such metrics we describe conditions on their dynamical evolution required to maintain asymptotic flatness. Pathologies are also identified in the evaporating white hole geometry in the form of an intermediate singularity accessible by timelike observers. We also describe a generalization of the equivalence between Rindler and Schwarzschild horizons to Kerr--Vaidya black holes, and describe the relevant geometric constructions.}

\end{abstract}

\maketitle

\section{Introduction}
Astrophysical black holes (ABHs) are dark, massive objects compact enough to possess a light ring. In the Milky Way alone, their population is estimated to be in the hundreds of millions. Models that attempt to describe ABHs typically fall into one of two distinct groups: those with and those without horizons. The  most common defining features of the former are the event horizon, a null surface that causally disconnects the black hole interior from the outside world, and the singularity, where curvature scalars diverge and the validity of general relativity breaks down \cite{CP:19,BCNS:19}. Solutions with an event horizon and singularity are referred as mathematical black holes (MBHs) \cite{F:14,MMT:22}. They have long been used as de facto proxies for studying black holes in both astrophysical and purely theoretical settings \cite{HE:73,W:84,C:92,FN:98,SKMHH:03,O:95,vF:15} and are consistent with all current observations.  Despite this, the notion that ABHs can be identified with MBHs remains speculation rather than  observationally established fact \cite{BCNS:19,M:23}.

Since the event horizon is unobservable in principle and there are many reasons to doubt its formation \cite{AB:05,A:20},  it is useful instead to focus on the most important feature of a black hole --- the trapping of light. The apparent horizon is the (foliation-dependent) boundary of the trapped region, and the trapped region itself is identified as the physical black hole (PBH) \cite{F:14,MMT:22}. In this setting, the occurrence of other features such as singularities is not assumed.

While black holes may be “the most perfect macroscopic objects in the universe,” with paradigmatic Schwarzschild and Kerr solutions exhibiting a wide range of useful symmetries \cite{C:92}, the complete picture involves complex astrophysical phenomena, including accreting matter and its back-reaction on the geometry, as well as Hawking radiation \cite{FN:98}. Consequently, analytical studies rely on simplified models, whose features depend heavily on the specific assumptions of the chosen framework.

The very discussion of ABH nature is necessarily framed using the classical geometric language of semiclassical gravity. Observational signatures of ABH models—such as gravitational waves, test particle motion, light rings, or black hole shadows—are assumed to follow classical general relativity or modified gravity theories.

 Semiclassical gravity \cite{W:84,F:05,K:12,MMT:22} provides the simplest theoretical framework that describes this classical geometry along with (potentially quantum mechanical) matter. Within this framework the Einstein equations balance the classical Einstein tensor with the renormalized energy-momentum tensor (EMT) $T_{\mu\nu}\defeq\6\hat T_{\mu\nu}\9_\omega$,
\be
G_{\mu\nu}\defeq R_{\mu\nu}-\half \sg_{\mu\nu}R=8\pi T_{\mu\nu}\ , \label{EE}
\ee
where $R_{\mu\nu}$ and $R$ are the Ricci tensor and scalar, respectively.
 The EMT possibly includes matter fields, terms arising from renormalization, and/or dark energy. In our approach we do not make any assumptions regarding the state $\omega$. We focus primarily on the Einstein tensor $G_{\mu\nu}$ and the
metric $g_{\mu\nu}$, with the EMT playing only an auxiliary role \cite{MMT:22,DSST:24}.

In spherical symmetry the assumption that a trapped region forms in finite time according to a distant observer   and that its boundary is singularity-free is sufficient to fully classify the near-horizon geometry and obtain physical properties of the resulting PBHs \cite{MMT:22,DSST:24,MMT:22D,DMT:22}.  Appendix~\ref{rev-spher} provides a brief review of PBHs and describes some interesting properties of their white hole counterparts.

However, ABHs are not expected to be spherically symmetric. Leaving aside the questions of efficiency of angular momentum transport and differences between single and binary black holes \cite{V:16,J:18,R:19,AGLT:24}, it is clear that even at the level of stylised models one should consider dynamical axially-symmetric systems. Moving to axial symmetry introduces a number of challenges due to the reduced symmetry \cite{HE:73,C:92}, while also leading to a host of important effects not present in the spherical case.

 In this  paper  we analyze the simplest dynamical axially symmetric models --- generalizations of the Kerr metric with variable mass $M$ and constant angular momentum to mass ratio $a\defeq J/M$, as well as those with variable $a$. Such metrics were introduced a long time ago \cite{MT:70,CK:77,CGK:90} and many of their properties are well understood \cite{X:99,ST:15,DT:20}.  Our goals are twofold. First, we present new features of these models and examine their relationship to spherically symmetric PBHs, identifying a weak singularity and an intermediate singularity within the near-horizon regions of certain models. Second, we discuss the equivalence (in the near-horizon approximation) between the analogs of the Rindler and Schwarzschild horizons in the axially symmetric case. While this equivalence is rigorously established for Rindler and Schwarzschild spacetimes, it is generally believed to hold more broadly \cite{P:10, P:15}. Recently, this relationship has been precisely formulated in Ref.~\cite{DS:23} using the concept of a separatrix \cite{F:14,york1983}. We extend this construction to Kerr–Vaidya black holes with both constant and variable angular momentum-to-mass ratios.

The rest of the paper is organized as follows. In Sec. \ref{sIII} we describe Kerr--Vaidya solutions with constant angular momentum to mass ratio, and classify them according to their EMTs, regularity, and horizon properties. The first two subsections briefly review some important and known properties of the solutions. In what remains we examine the properties and motion of test particles near the horizon of these black holes. In Sec. \ref{sIV} we consider generalizations of Kerr--Vaidya metrics with variable angular momentum to mass ratios.  In Sec.~\ref{sV} we discuss how the near-horizon description of the Schwarzschild black hole in terms of the Rindler metric generalizes to Kerr--Vaidya black holes. We conclude in Sec~\ref{discus} with a summary of our results, their implications, and directions for future work. Throughout, we work in units where $\hbar=c=G=1$ and use the $(-+++)$ signature.

\section{Kerr--Vaidya spacetimes} \label{sIII}

  Kerr--Vaidya metrics are the simplest non-stationary generalizations of the Kerr metric.  Throughout this work, we will distinguish between two distinct classes of metrics which may be reasonably called Kerr-Vaidya. The first are those with variable mass $M$ and constant $a=J/M$. This class of metrics represents the simplest possible generalization of the Vaidya metric to axially symmetric form, and contains the Vaidya-Patel metric as an example \cite{VP:73}. These we refer to simply as {\it Kerr-Vaidya} (KV) where there is no ambiguity. The second class of metrics we examine are those with both variable $M$ and variable $a$. We refer to this more general class of metrics as {\it generalized Kerr-Vaidya} (gKV) throughout. We admit the possibility that other metrics exist which may also be reasonably termed Kerr-Vaidya (or some variation thereof) and note that our analysis is restricted only to those metrics explicitly defined below. In this Section we are only concerned with KV metrics, while the analysis of what we term gKV metrics is presented in Sec. \ref{sIV}.

  The easiest formal way to obtain the Kerr metric is to follow the complex-valued  Newman--Janis  transformation \cite{C:92,SKMHH:03} starting with the Schwarzschild metric written in either retarded or advanced Eddington--Finkelstein coordinates. Kerr--Vaidya metrics result from the Kerr metric if the mass is instead allowed to be function of the advanced or retarded null coordinate \cite{heje:79,DT:20}. These can also be obtained through the application of a Newman-Janis transformation to the corresponding Vaidya metrics~\cite{CK:77, DT:20}, and can be generalized to gKV metrics by allowing $a$ to also vary as a function of a null coordinate. Hence a practical starting point is the Kerr metric written using the ingoing or outgoing principal null congruences \cite{C:92,O:95} and making the mass dependent on the relevant null coordinate.

  The advanced/ingoing Kerr--Vaidya metric is given by
\begin{align}
 ds^2=&-\bigg(1-\frac{2 M r}{\rho^2}\bigg)dv^2+2 dv dr- \frac{4 a M r \sin^2\theta}{\rho^2} dvd\psi \nonumber \\
& -2 a \sin^2\theta dr d\psi  \!  +\rho^2 d\theta^2 + \frac{\Sigma^2}{\rho^2}\sin^2\theta d\psi^2\ ,
    \label{kv}
\end{align}
where  $M=M(v)$,  while the retarded/outgoing Kerr--Vaidya metric is given by
  \begin{align}
      ds^2=&-\bigg(1-\frac{2 M r}{\rho^2}\bigg)du^2-2 du dr -      \frac{4 a M r \sin^2\theta}{\rho^2}du d\psi   \nonumber \\
   & +2 a \sin^2\theta d\psi dr \!   +\rho^2 d\theta^2   +
    \frac{\Sigma^2}{\rho^2}\sin^2\theta d\psi^2\ ,
    \label{ku}
    \end{align}
where  $M=M(u)$. In the above,
\begin{align}
&\rho^2\defeq r^2+a^2 \cos^2\theta\ , \\
&\Delta\defeq r^2-2 M r+a^2\ , \\
&\Sigma^2\defeq(r^2+a^2)^2\!-\!a^2 \Delta \sin^2\theta\ ,
\end{align}
and $a=J/M$ is the angular momentum per unit mass. For the stationary Kerr metric the null coordinates   are given by
\be
\begin{aligned}
du_+&\equiv dv=dt+\frac{r^2+a^2}{\Delta}dr\ , \\
 du_-&\equiv du=dt-\frac{r^2+a^2}{\Delta}dr\ ,
\end{aligned}
\ee
and
\be
 d\psi_\pm=d\phi\pm \frac{a}{\Delta}dr\ ,
\ee
where $\phi$ is the usual (Boyer--Lindquist) azimuthal angle. In the following, we omit the subscript on the variable $\psi$ as it does not lead to confusion. For Kerr--Vaidya metrics the integrating factors have to be introduced.

One important observation should be made at this stage: of the four seed Vaidya solutions that are used to generate the Kerr--Vaidya metrics, only two  --- a PBH with $r_\sg'(t)\equiv r_+'(v)<0$ and a white hole with $r_\sg'(t)\equiv r_-'(u)>0$ --- can be realized if one requires finite formation time according to a distant observer. All four types of metrics described by Eqs.~\eqref{kv} and \eqref{ku} violate the null energy condition (NEC) \cite{HE:73,F:05,SKMHH:03}, and are thus compatible with the finite formation time of the objects they describe \cite{HE:73,FN:98,MMT:22}. Moreover, test particle trajectories in this case have number of interesting peculiarities (see Ref.~\cite{DMT:22} and Appendix \ref{apW} for details).

We first test the applicability of the above solutions to black or white holes by confirming a number of their intuitive properties, and then using the decomposition of Ref.~\cite{MT:70} perform a complete classification of their EMTs.  We derive the equations of motion for test particles in these Kerr--Vaidya backgrounds, and show that their trajectories have qualitatively different features compared to their motion in the Kerr background, but have similarities with the dynamics of test particles on the background of the admissible Vaidya solutions. Finally, we show that metrics \eqref{ku} with $M'(u)<0$ develop an intermediate singularity as $M(u)\to a$.

\subsection{Classification of solutions}

 We first review some essential properties of Kerr and Kerr-Vaidya solutions. The event horizon of a Kerr black hole is located at
\be
  r_0\defeq M+\sqrt{M^2-a^2}\ ,
  \ee
which is the largest root of the equation $\Delta=0$. Compact surfaces of constant $u$ and $r$ (or $v$ and $r$) are of spherical topology. The introduction of two families of null geodesics $l^\mu_\pm$ normal to these spheres allows one to identify the domain $\Delta<0$ as a black hole in $(v,r)$ coordinates (i.e. both expansions of the null congruences are negative, $\vartheta_\pm<0$), and the same domain in $(u,r)$ coordinates as a white hole (i.e. both expansions satisfy  $\vartheta_\pm>0$).

It is important to distinguish between the black and white hole Kerr--Vaidya solutions. 
In fact, there is a certain confusion in the literature, including Ref.~\cite{DT:20} of some of the present authors. The construction of the two families of null geodesics (for example, as introduced in Ref.~\cite{FT:08}) indicates that for $r<r_0$ the  expansions of both families of congruences are positive for the Kerr--Vaidya metrics in $(u,r)$. This suggests that these are models of white holes. The metrics in $(v,r)$ coordinates instead describe black holes.

Unlike the Vaidya metrics, both growing and contracting Kerr--Vaidya black and white hole solutions violate the NEC. While occasionally treated as a telltale sign that the solutions are non-physical, NEC violation actually implies that these metrics describe objects which may form in finite time according to distant observers. For reference we summarise here some properties of their EMTs.

Using the null vector $k^\mu=(0,1,0,0)$ the EMT of both the advanced and retarded Kerr--Vaidya metrics can be written concisely \cite{MT:70,DT:20} as
\be
T_{\mu\nu}=T_{oo}k_\mu k_\nu+q_\mu k_\nu+q_\nu k_\mu\ , \label{MT-emt}
\ee
where $T_{oo}$ stands for either $T_{uu}$ or $T_{vv}$ and the components of $T_{\mu\nu}$ and of the auxiliary vector $q_\mu$ (which satisfies $q_\mu k^\mu=0$) are given in Appendix~\ref{aB}.

The EMTs are characterized by the Lorentz-invariant eigenvalues of the matrix $T^{\ha}_{~~\hat{b}}$, i.e.,
 the roots of the equation
\be
 \det (T^{\ha\hat{b}}-\lambda \eta^{\ha\hat{b}})=0\ , \qquad \eta^{\ha\hat{b}}=\mathrm{diag}(-1,1,1,1)\ .
\ee
Using a tetrad in which the null eigenvector $k^\mu=k^{\ha}e^\mu_\ha$ has the components $k^\ha=(1,1,0,0)$,
the third vector $e_{\hat 2}\propto \pad_\theta$ and the remaining vector $e_{\hat{3}}$ is found by completing the basis. The EMT then takes the form
 \be
T^{\ha\hat{b}}=\left(\begin{tabular}{cc|cc}
 $\nu$ & $-\nu$  & $-q^{\hat 2}$& $-q^{\hat 3}$\\
 $-\nu$ & $\nu$  & $q^{\hat 2}$& $q^{\hat 3}$\\ \hline \label{IVP}
 $-q^{\hat 2}$& $q^{\hat 2}$ & 0 & 0 \\
$-q^{\hat 3}$ & $q^{\hat 3}$& 0 & 0
 \end{tabular}\right)\ .
\ee
Explicit expressions for  the tetrad vectors and the EMT matrix elements are given in Appendix \ref{aB}.

The EMT of Eq.~\eqref{IVP} has a four-fold degenerate Lorentz-invariant eigenvalue $\lambda=0$. The two non-zero eigenvectors corresponding to this eigenvalue are
\begin{equation}
    (1,1,0,0), \quad \big(0,0,1,-q^{\hat 2}/q^{\hat 3}\big)\ ,
\end{equation}
which are null and spacelike, respectively. Thus for $a\neq 0$ the EMT is of type III in the Hawking--Ellis classification \cite{HE:73} (or type $[(1,3)]$ in a more refined  Segre classification \cite{SKMHH:03}), indicating that the NEC is violated as the eigenvectors are triple null. Note that the amount of allowed NEC violation is bounded in quantum field theory (on a curved background) by various quantum energy inequalities \cite{KS:20}, though these bounds will not play an important role in the scenarios described here.

   \subsection{Horizons}

 {We recall some basic facts concerning horizons in Kerr and Vaidya spacetimes before presenting our semi-classical analysis.} The apparent horizon of the Kerr black hole coincides with its event horizon, which is a null surface. For  both the ingoing and  outgoing  Vaidya metrics the apparent/anti-trapping horizon is located at $r_\sg=2M$, and is timelike for both admissible types of solutions ($r_+'(v)<0$ and $r_-'(u)>0$). The situation is more involved for the Kerr--Vaidya metrics.

  For the  metric \eqref{ku} the relation $r_-=r_0$ also holds \cite{X:99}. If this hypersurface is represented as $\Phi(u,r)=r-r_0(u)$, then the normal vector $\pad_\mu\Phi$ satisfies
  \be
  \Phi_\mu\Phi^\mu|_{r=r_0} =-\frac{4 r_{0}^{2} M'\left(r_{0}^{4}-a^4+r_{0}^{2} a^2 \sin^2\theta M'\right)}{(r_{0}^{2}+a^2 \cos^2\theta)\,(r_{0}^{2}-a^2)^2}\ .
  \ee
As a result, the anti-trapping horizon of a shrinking white hole is spacelike. For  expanding white holes it is timelike so long as the black hole is not too close to being extreme ($a\approx M$) and its growth rate is not too large.

This is not so for the black hole metrics of Eq.~\eqref{kv} \cite{ST:15}, where the expansion of the outgoing null congruence at $r=r_0$ is $\vartheta_+|_{r_0}= M'r_0 a^2\sin^2\theta/\big(4(r_0^2+a^2)^2\big)$. To leading order in $M'$ the location of the apparent horizon can be expressed as
\be
r_+=r_0(v)+M'(v)\varsigma(r_0,\theta)\ , \label{ah-kv}
\ee
$\varsigma\leqslant 0$, and can be obtained numerically \cite{DT:20}.

Following the same steps for $\Phi(v,r,\theta)=r-r_+(v,\theta)$ we find that the normal vector satisfies
\be
  \Phi_\mu\Phi^\mu|_{r=r_0} =-\frac{2 M'\big(2M r_0^2-(r_0-M)^2\varsigma\big)}{(r_{0}^{2}+a^2 \cos^2\theta)( r_{0}-M)} +\cO(M'^2, M'')\ .
  \ee
Hence for a slowly evaporating/accreting black hole the apparent horizon is timelike/spacelike. The same is true for the hypersurface $\Delta(r,v)=0$.

\subsection{Admissible solutions}

We now discuss the compatibility of the Kerr--Vaidya solutions with the requirements of regularity and finite formation time and its consequences. We investigate separately the black hole and the white hole Kerr--Vaidya solutions. In the proper frames of test particles falling into spherically-symmetric PBHs and/or their counterpart white holes, the values of (negative) energy density and pressure are finite at the horizon. Mild firewalls (hypersurfaces where these quantities are divergent) are possible only for non-geodesic observers with monotonically increasing four-acceleration. In spherical symmetry, a radially infalling Alice is a zero angular momentum observer (ZAMO) \cite{FN:98}.  However in axially symmetric spacetimes, the requirement of zero angular momentum along the axis of rotation ($L_z=0$)  results in a non-trivial angular velocity $\Psi_Z$ of Alice via the condition $L_z\defeq\xi_\psi \cdot u_\mathrm{A}=0$, where  the Killing vector is $\xi_\psi=\pad_\psi$. In what follows we consider infalling ZAMO observers. While the NEC is violated for all four classes of the Kerr--Vaidya solutions, an ingoing Alice measures a positive energy density near the horizon of a growing black hole and evaporating white hole. The spacetime of evaporating Kerr--Vaidya white hole exhibits timelike geodesic incompleteness.

\subsubsection{Black hole solutions}

For the black hole solutions (Kerr--Vaidya metrics in $(v,r)$ coordinates) Alice's four-velocity is
\be
u_\rA=(\dot v, \dot r, \dot \theta,\dot \psi_Z)\ , \label{zamo49}
\ee
where the ZAMO condition implies that $\dot\psi_Z=-(\sg_{v\psi}\dot v+\sg_{r\psi}\dot r)/\sg_{\psi\psi}$. The normalization of the four-velocity   results in
\be
\dot v=\frac{1}{\Delta}\left((r^2+a^2)\dot r\pm\Sigma\sqrt{\Delta \big(\rho^{-2}\delta+\dot\theta^2\big)+\dot r^2}\right)\ , \label{dV}
\ee
where $\delta=0,1$ for null and timelike test particles, respectively.
 The choice of physically relevant solution is determined not by whether Alice is inside or outside the trapped region, but rather by her position relative to the domain  $\Delta(v,r)<0$. If $\Delta>0$, then both ingoing and outgoing trajectories  correspond to the upper sign in \eqref{dV}. This can be seen by  comparison with the Kerr metric using the explicit transformation $v=v(t,r)$, as well as by taking the limit $a\to 0$ and comparing directly with the Vaidya metric. 

It is possible to have $\dot v<0$ if
\be
 \dot r<0, \qquad \dot r^2>\dot r_\mathrm{zero}^2\defeq \frac{\Sigma^2(\rho^{-2}\delta+\dot\theta^2)}{a^2\sin^2\theta}\ . \label{vrsmall}
\ee
  For comparison, the tangents to the ingoing and the outgoing  principal null congruence are
\be
k_-^\mu=(0,-1,0,0), \qquad k_+^\mu=(r^2+a^2,\half\Delta, 0,a)/\rho^2\ ,
\ee
respectively, with $k_+^\mu k_{-\mu}=-1$.

 For $\Delta<0$ it is necessary to have $\dot r^2\geqslant-\Delta(\rho^{-2}\delta+\dot\theta^2)$. For the Kerr metric the time orientation for $r<r_0$ (in the so-called region II) is established \cite{O:95} by using the future-directed null vector field $k^\mu_-$. As a consequence only causal trajectories with $\dot r<0$ are admissible, and this extends to the Kerr--Vaidya metric as $M$ is assumed to be a sufficiently smooth function of $v$.  The same conclusion is reached if one requires consistency with the Vaidya metric solutions in the limit $a\to 0$. The upper sign in  Eq.~\eqref{dV} corresponds to the outgoing and the lower sign to the ingoing geodesic.

Close to $\Delta=0$ for ingoing Alice, we find that
\be
\dot v=\frac{(r^2+a^2)(1+\rho^2\dot\theta^2)}{2|\dot r|\rho^2}-\frac{|\dot r|a^2\sin^2\theta}{2(r^2+a^2)} +\cO(\Delta)\ , \label{vrv}
\ee
if $\dot r^2\gg \Delta/\rho^2$, and thus $\dot v\sim \dot r a^2/r_0^2$ when $\dot r\to -\infty$.  If instead $\dot r^2\lesssim\Delta/\rho^2$, then
\be
\dot v=\frac{r^2+a^2}{\rho\sqrt{\Delta}}\left(\sqrt{1+\mfr^2 +\rho^{2}\dot\theta^2}-\mfr\right)+\cO(\sqrt{\Delta})\ ,
\ee
where $0\leqslant\mfr^2\defeq\lim \dot r^2\rho^2/\Delta$ as $r\to r_0$.

The most efficient way to study the trajectories of test particles on the Kerr background is by using the Hamilton--Jacobi equation \cite{C:92,FN:98}, as it allows for a complete separation of variables. However, in Kerr--Vaidya spacetimes energy is not conserved, and we instead deal directly with the geodesic equations. We represent the second derivatives as
\begin{align}
&\ddot r+\eD_r=-\frac{2rM'}{\Sigma^2}\big[(r^2+a^2)\dot v^2-a^2\sin^2\!\theta\, \dot v\dot r\big] \nonumber \\
&\qquad \quad \quad -(a^2+r^2) \pad_v L/\rho^2\ , \label{vddr} \\
&\ddot v+\eD_v=-a^2\sin^2  \pad_v L/\rho^2\ , \\
&\ddot \theta +\eD_\theta=0\ ,
\end{align}
where $\eD_r$, $\eD_v$, and $\eD_\theta$ contain all of the terms that appear  in the $M=\mathrm{const.}$ case and can be read from Eq.~\eqref{EOMv}, and
\be
\pad_v L=-\frac{r\rho^2M'}{\Sigma^4}\big[(r^2+a^2)\dot v-a^2\sin^2\!\theta\dot r\big]^2.
\ee
More  details are given Appendix \ref{appd}. We see that the additional terms $\eD_r$, $\eD_v$, and $\eD_\theta$ are regular at $r=r_0(v)$.  Note that for $a=0$ the right hand side of Eq.~\eqref{vddr} reduces to its Vaidya counterpart $-M'(v)\dot v^2/r$.

For slow particles, Eqs.~\eqref{vrv} and \eqref{vddr}  implies that the radial acceleration diverges as
\be
\ddot r=-\frac{2M' r}{\Delta}\big(r^2+a^2\big)\big(\rho^{-2}+\dot \theta^2)\ .
\ee

For an evaporating black hole, i.e. $M'(v)<0$,   the apparent horizon lies outside of the hypersurface $r=r_0(v)$. As in the spherically-symmetric case \cite{MMT:22}, a sufficiently slow ingoing test particle stops at some $\tilde{r}>r_0 $ and then moves (at least temporarily) on an outgoing trajectory. Otherwise it crosses the hypersurface $r=r_0(v)$ with all components of the four-velocity being finite. In the latter case the proper energy density $\varrho_A$ is negative and finite, and its explicit expression for the motion in the
equatorial plane ($\theta=\pi/2$, $\dot \theta=0$) is
\be
\varrho= \frac{T_{vv}}{4\dot r^2}+\cO(a^2)\ .
\ee

For a growing black hole we have that $M'(v)>0$, and thus the apparent horizon is located inside the hypersurface $r=r_0(v)$. Additional terms act to decelerate the ingoing particle but it remains finite.

As with PBHs, slow particles initially falling into an evaporating Kerr--Vaidya black hole can stop and reverse direction, which occurs at some $r_\#>r_0$. For a growing black hole the infall of slow particles can only accelerate relative to the Kerr--Vaidya metric, while the effect is insignificant for fast particles.  Hence, the energy density in Alice's frame remains finite throughout the entire infall.

On the other hand, it is possible to attempt to force the test particle to cross the surface $\Delta=0$ of an accreting black hole on  a non-geodesic trajectory with a nearly zero (or even positive) value of $\dot r$. In the latter case
\be
\dot v\approx\frac{2(r^2+a^2)\dot r}{\Delta}\ ,
\ee
near $\Delta=0$, and the energy density diverges as $\Delta^{-2}$,
\begin{align}
	\rho_\rA &\approx\bigg(T_{v v}+ 2T_{v \psi}\frac{a}{r^2+a^2}+ T_{\psi\psi}\frac{a^2}{(r^2+a^2)^2}\bigg)\dot{v}^2 \nonumber \\
    & \approx \frac{4 (2 r M_v- a^2 \sin^2\theta M_{v v}) r}{8\pi \Delta^2} \dot{r}^2\ .
\end{align}
This divergence  indicates that $\Delta=0$ is a weakly singular surface.

Finally, we observe the following asymmetry between the accreting and evaporating stages of the evolution of Kerr--Vaidya black holes.  During  accretion, as indicated by Eq.~\eqref{ah-kv},  the apparent horizon lies inside the hypersurface $r=r_0(u)$ and the expansion of at least some families of outgoing null geodesics is positive. Nevertheless, no signalling to distant observers is possible until the growth stops and the evaporation begins.

\subsubsection{White hole solutions}\label{kvwh}

For white hole solutions the ZAMO Alice has four-velocity
   \be
 u^\mu_\mathrm{A}=\big(\dot{u},  {\dot{r}},  \dot{\theta},\dot \psi_Z\big)\ , \label{zamo63}
 \ee
where  $\dot\psi_Z=-(\sg_{u\psi}\dot u+\sg_{r\psi}\dot r)/\sg_{\psi\psi}$.  The normalization of the four-velocity together with the ZAMO condition  results in
\be
\dot u=\frac{1}{\Delta}\left(-(r^2+a^2)\dot r\pm\Sigma\sqrt{\Delta\big(\rho^{-2}\delta+\dot\theta^2\big)+\dot r^2}\right)\ . \label{dV2}
\ee
The choice of physically relevant solution is again determined by the position of Alice relative to the domain  $\Delta<0$. If $\Delta(r)>0$, then both the ingoing and outgoing trajectories  correspond to the upper sign, and this includes a transition from   an outgoing to an ingoing trajectory.

Close to $\Delta=0$ for an ingoing Alice
\be
\dot u=\frac{2|\dot r|}{\Delta}(r^2+a^2)+\cO\big(\Delta^0\big)\ , \label{dudiv}
\ee
when $\dot r^2\gg \Delta/\rho^2$,  and
\be
\dot u=\frac{r^2+a^2}{\rho\sqrt{\Delta}}\left(\sqrt{1+\mfr^2 +\rho^{2}\dot\theta^2}+\mfr\right)+\cO(\sqrt{\Delta})\ .
\ee
when $\dot r^2\sim\Delta/\rho^2$, where $0\leqslant\mfr^2\defeq\lim \dot r^2\rho^2/\Delta$ as $r\to r_0$. Finally, for  $\dot r>0$ the rate of change of the retarded coordinate is finite,
\be
\dot u=\frac{(r^2+a^2)(1+\rho^2\dot\theta^2)}{2|\dot r|\rho^2}-\frac{\dot ra^2\sin^2\Theta}{2(r^2+a^2)} +\cO(\Delta)\ . \label{outdU}
\ee

The null outgoing congruence with the tangent $l_+^\mu=(0,1,0,0)$ that defines the $(u,r)$ coordinates in for the Kerr metric remains geodesic for $M'(u)\neq 0$, as can be easily seen from the geodesic equations, with $r$ being the affine parameter.

The geodesic equation implies that
\begin{align}
&\ddot r+\eD_r=\frac{2rM'}{\Sigma^2}\big[(r^2+a^2)\dot u^2+a^2\sin^2\!\theta\, \dot u\dot r\big] \nonumber \\
&\qquad \quad \quad +(a^2+r^2) \pad_u L/\rho^2\ , \label{uddr} \\
&\ddot u+\eD_u=a^2\sin^2  \pad_u L/\rho^2 \\
&\ddot \theta +\eD_\theta=0\ ,
\end{align}
where $\eD_r$, $\eD_u$, and $\eD_\theta$ contain all of the terms that appear  in the $M=\mathrm{const.}$ case and
\be
\pad_u L=-\frac{r\rho^2M'}{\Sigma^4}\big[(r^2+a^2)\dot u+a^2\sin^2\!\theta\dot r\big]^2\ .
\ee
 The details are given Appendix \ref{appd}. For $a=0$ the right hand side of Eq.~\eqref{uddr} reduces to its Vaidya counterpart of Eq.~\eqref{eqR}.

 For an ingoing particle with $\dot r^2\gg\Delta/\rho^2$ and $\Delta\ll r_0^2$ we have
 \be
 \ddot r=\frac{r M'\dot u^2}{r^2+a^2}+\cO\big(\Delta^{-1}\big)=\frac{r M' (r^2+a^2)\dot r^2}{(r-M)^2 z^2} + \cO\big(z^{-1}\big)\ , \label{ddruA}
 \ee
 where we introduced the gap function  $z(\tau)\defeq r(\tau)-r_0\big(u(\tau)\big)$.

As in the case of PWHs, Eq.~\eqref{ddruA}  indicates that particles initially falling into a growing  Kerr--Vaidya white hole are stopped and reversed at some radius $\tilde{r}>r_0$. If the particle is later overtaken by an expanding anti-trapping horizon the proper energy density $\varrho_\rA=T_{\mu\nu}u_\rA^\mu u_\rA^\nu$ is finite.

On the other hand, in the spacetime of a sufficiently slowly contracting Kerr--Vaidya white hole the local energy density as measured by infalling Alice can reach arbitrarily high values.
First we note that as $\dot z=\dot r -r_0'\dot u$, for a given $\dot r(\tau)$ and $\dot u(\tau)$ during the evaporation (i.e., while $M>a$) the gap decreases only if the gap separatrix at
\be
z_*\approx|r'_0|\frac{r^2+a^2}{\sqrt{M^2-a^2}}\ ,
\ee
is not crossed. As a result for $z\approx z_*$
\be
\ddot r=-\frac{(r-M)^2\dot r^2}{r(r^2+a^2)|M'|} +\cO(z^{-1})\ .
\ee
The radial acceleration $\ddot r$ can be bounded analogously to Appendix \ref{apW}. The local energy density for an infalling Alice   $\varrho_\rA\approx T_{uu}\dot u^2$ during the scales as (Appendix~\ref{appd} provides more computational details),
\be
\varrho_\rA\sim\frac{\dot r^2|M'|}{z^2_*}\propto\frac{1}{|M'|} . \label{KV-rd}
\ee
Hence in the contracting Kerr--Vaidya white hole solution  the timelike geodesics may reach the intermediate singularity (also known as a matter singularity) outside the event horizon.

\begin{figure}[!htbp]
	\includegraphics[width=\linewidth]{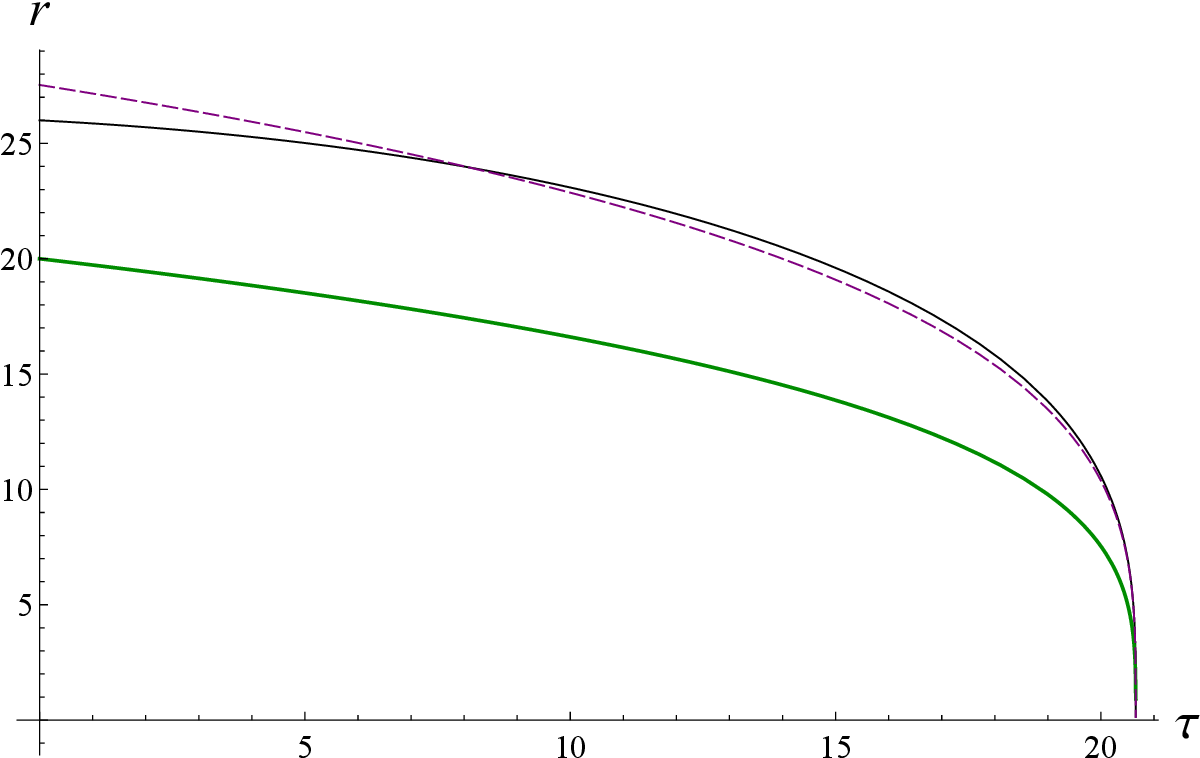}
	\caption{Timelike ZAMO geodesics on the equatorial plane ($\theta=\frac{\pi}{2}$) of retarded Kerr-Vaidya spacetime. The thin black line is a timelike geodesic. The solid green line and thin dashed purple line are the apparent horizon and the gap seperatrix, respectively. The mass evolves as $m(u)=10-0.05 \,u$ and the angular momentum per mass is $a=0.1$. The initial conditions are $r(\tau=0)=26$, $u(\tau=0)=0$, $\phi(\tau=0)=0$, $\dot{r}(\tau=0)=-0.12.$}
	\label{fig:exampleKV}
\end{figure}

\section{Kerr--Vaidya solutions with variable $a$} \label{sIV}

An important generalization of the Kerr--Vaidya solution which we now consider involves an angular momentum to mass ratio which varies a function of time. While such gKV metrics present some difficulties (such as having no Kerr-Schild form \cite{CGK:90}), it is nonetheless an important generalization to consider for a number of reasons.  {For one, particle emission through Hawking radiation (which is often modelled by the Vaidya metric in the absence of rotation) forces one to contend with variable $a$. Though the detailed balance between angular momentum and mass loss through Hawking radiation depends on the spectrum of emitted particles and their angular momentum modes, the emission process is nonetheless dominated by massless excitations (photons and gravitons) for most of the lifetime of astrophysically sized black holes. For such emission, it is known that $a^*=J/M^2$ tends to zero \cite{P:76,DKS:16,AAS:20}, implying that $J\rightarrow0$ faster than $M^2\rightarrow0$ and thus $a=J/M$ cannot be constant. This will be the case when $T_H\sim M^{-1}\leqslant m_{\text{min}}$, where $m_{\text{min}}$ is the mass of the lightest known particle. An upper bound on the neutrino mass of $m\leqslant3.52\times 10^{-30}$ (in Planck units and averaged over three flavours \cite{PDG:22}) then means this will be the case for black holes of mass $M\geqslant 2.84\times 10^{29}=6.18\times10^{24}$ g. For reference, a solar mass black hole has $M_{\odot}=2\times10^{33}$ g.} The assumption that $a=J/M=\mathrm {const.}$ would also be incompatible with the continuous eventual evaporation of a physical black hole, since for $M <a$ the equation $\Delta=0$ has no real roots and the Hawking temperature
\be
T=\frac{1}{2\pi}\left(\frac{r_0-M}{r_0^2+a^2}\right)\ ,
\ee
which is proportional to the surface gravity, goes to zero as $M\to a$.

gKV metrics may also be of relevance for certain astrophysical scenarios where the shedding of angular momentum can occur sufficiently rapidly to require $a\neq \text{const.}$.  This can occur both through variants of the Penrose process \cite{DTAS:18} where interaction with the accretion disk can remove angular momentum with enormous efficiency, as well as generic interactions between accreting matter and the black hole \cite{AF:13}.  A generic feature of both scenarios is a complicated set of angular momentum transport phenomena between the matter and the black hole. In such cases a precise matching of the rate of angular momentum flux $\dot{J}$ and mass growth $\dot{M}$ of the black hole would represent a highly fine-tuned scenario. We therefore expect that generically, situations which do admit a useful description in terms of an axially symmetric radiating metric will generically require considering gKV metrics ($a\neq$ const.).

Geometries with variable $M$ and $a$ are considerably more intricate. The Newman--Penrose formalism \cite{C:92,SKMHH:03} is based using a null tetrad, which consists of two real null vectors (such as $k_\pm^\mu=(e_{\hat 0}^\mu\pm e_{\hat 1}^\mu)/\sqrt{2}$ that are constructed from the vectors given in Appendix \ref{aB}), and two complex conjugate null vectors that are constructed from a pair of real orthonormal spacelike  vectors that are also orthogonal to the vectors $k^\mu_\pm$.   The ten independent components of the Weyl tensor can be represented by five complex scalars, denoted as $\Psi_0, \ldots,  \Psi_4$. Subjecting the null tetrad to some Lorentz transformation it is possible to have
\begin{equation}
 \Psi_0\to\tilde{\Psi}_0 =  \Psi_0+ 4 b \Psi_1+ 6 b^2 \Psi_2+ 4 b^3 \Psi_3+ b^4 \Psi_4\ ,
\end{equation}
where $b$ is a complex scalar. The roots of the equation $\tilde\Psi_0=0$ play a crucial role in the algebraic classification of spacetimes. If the equation admits four distinct roots, the spacetime is classified as algebraically general; otherwise, it is algebraically special.   {Kerr--Vaidya metrics with $a=\mathrm{const.}$ are of Petrov type II unlike their Kerr counterparts, which are of Petrov type D. This is why these metrics posses  principal null geodesic congruences and can  be written in Kerr--Schild form. However, the gKV metrics (with variable $a$) are of Petrov type I (algebraically general) and thus cannot be cast into Kerr-Schild form \cite{C:92,CGK:90}. Thus the question of their asymptotic flatness is quite involved. To investigate the asymptotic flatness of the gKV solutions, we explicitly calculate components of the Riemann tensor in the Newman--Penrose formalism and determine their asymptotic behavior, checking whether each component falls off sufficiently rapidly to ensure asymptotic flatness in the strict sense of Bondi and Sachs \cite{B:60,S:62}. {We find that the future asymptotic flatness for gKV metrics (those with $a\neq$ const.) depends on the rate of change of $a=J/M$}, requiring that $a\rightarrow$ const. at least as fast as $|a'|\propto 1/v$. We provide more details in Appendix \ref{aB-var}.}

The EMT $T^{\ha}_{~~\hat{b}}$, written in an orthonormal basis, has four distinct Lorentz-invariant eigenvalues. Two eigenvalues are complex, and two are real, corresponding to spacelike eigenvectors. This observation implies that the EMT is type IV (see Appendix \ref{aB-var} for details).

Variable $a$ introduces additional terms to the geodesic equations. In the advanced metrics the additional terms are finite. In the retarded metrics, the (potentially) most divergent term contributing to the right hand side of Eq.~\eqref{uddr} is
\be
\delta \ddot r=\kappa aa'\dot u^2+\cO(\dot u)\ ,
\ee
where
\be
\kappa=\frac{2 r(a^2+r^2)M\left((a^2+r^2)\rho^2\cos^2\theta+2r^3M\sin^2\theta\right) }{\rho^6\Sigma^2}>0\ .
\ee
If the evaporation does not stop until $M=a=0$, the divergence of Eq.~\eqref{KV-rd} does not occur. However, as the signs of $a'$ and $M'$ coincide, the dynamics will be qualitatively similar to the Kerr--Vaidya case.

\section{Rindler Horizons for Kerr--Vaidya black holes}\label{sV}

The near-horizon geometry of a Schwarzschild black hole can be conformally mapped to a Rindler spacetime \cite{FN:98,BMPS:95,P:15}
	\begin{equation}
		ds^2= -\epsilon^2 dt^2+ d\epsilon^2+ dY^2+ dZ^2\ ,
	\end{equation}
	which has a horizon at $\epsilon=0$ (called the Rindler horizon) that separates the left and right Rindler wedge of Minkowski space. In the Rindler decomposition of Minkowski space, observers with constant proper acceleration $\mathfrak{a}$ follow orbits of the Lorentz boost operator $K$, and thus observe the Minkowski vacuum as a thermal density matrix at temperature $T_U=\mathfrak{a}/2\pi$ due to the presence of a Rindler horizon. In the Schwarzschild geometry, a local observer hovering at fixed areal radius near the event horizon requires a proper acceleration $\mathfrak{a}\sim1/4M$ with respect to the asymptotic frame in order to maintain their position. The corresponding Unruh temperature is then just $T_U=1/8\pi M$, which is precisely the Hawking temperature.

It is worth noting that despite this formal analogy, the Hawking and Unruh effects remain physically distinct. Spatial homogeneity of the Unruh temperature cannot be maintained for macroscopic objects \cite{B:15}, and the radiation is observed as an isotropic heat bath \cite{G:11}, while Hawking radiation is not. Even in the infinite mass limit (where one expects the Rindler approximation to the near-horizon region to become exact) the topological difference between the Schwarzschild and Rindler geometries enters as a difference in sub-leading factors for the entanglement entropy across the respective horizons \cite{sS:11}.

	Nonetheless, such a near-horizon Rindler approximation has considerable utility and has been used for computing black hole entropy \cite{C:95}, generalizing the laws of black hole mechanics \cite{B:13}, and in the study of asymptotic symmetries \cite{H:10}. A similar relationship is expected to hold in a more general dynamical setting \cite{P:15, MMT:22}, and has been shown to hold also for PBHs.

For the Vaidya black hole (a PBH with $w_1\approx 0$ in the notation of Sec. \ref{apW}) it has been shown that there is a precise conformal transformation between the near-horizon geometry the  Rindler space \cite{DS:23}, allowing one to compute the associated Hawking temperature in the dynamical background. The general feature of black hole geometries admitting a Rindler/conformal Rindler description near their horizons allows one to associate yet another horizon to a black hole---the Rindler horizon. In the examples we discuss here, the Rindler horizon of a black hole coincides with the separatrix (the outgoing radial (ZAMO) null curve for which $d^2r/d\tau^2 = 0$). For black holes evolving adiabatically the separatrix very closely approximates the event horizon, making it a useful notion of horizon in the discussion of dynamical black hole evaporation \cite{BMPS:95,MMT:22}.

\subsection{Case 1: constant $a$}

The Schwarzschild-Rindler analogy described above can readily be extended to the axially symmetric case. To determine the location of the separatrix for the Kerr--Vaidya metric with constant $a$ we consider an outgoing null curve, which to leading order is given by
\begin{equation}
	\dot r= \frac{\Delta}{2(r^2+ a^2)}\ , \label{eq10}
\end{equation}
where we have chosen a parametrization such that $\tau=v$. Differentiation with respect to $\tau$ gives
\begin{equation}
	\frac{d^2r}{d\tau^2}= \frac{(r\dot r- \dot M r- M\dot r)(r^2+ a^2)- \Delta r \dot r}{(r^2+a^2)^2}\ .
\end{equation}
Using \eqref{eq10} and imposing the condition that $\frac{d^2r}{d\tau^2}=0$ then gives
\begin{equation}
	\frac{d^2r}{d\tau^2}= \frac{(r\Delta- 2 M' r(r^2+ a^2)- M\Delta)(r^2+a^2)- \Delta^2r}{2(r^2+a^2)^3}= 0\ .
\end{equation}
Keeping only the leading order terms in the near horizon expansion implies that
\begin{equation}
	\Delta\approx \frac{2 M' r_0(r_0^2+a^2)}{r_0-M}\ . \label{sep9}
\end{equation}
On the other hand, the Kerr--Vaidya metric can be written in the form
\begin{equation}
	ds^2= -\frac{\Delta \rho^2\ d\tilde v^2}{(r^2+ a^2)^2} + \frac{2\rho^2\ d\tilde v dr}{r^2+ a^2} + \rho^2( d\theta^2+  \sin^2\theta d\tilde\phi^2)\ , \label{nc52}
\end{equation}
where
\be\begin{aligned} \label{nc53}
	d\tilde v&= \frac{(r^2+ a^2)}{\rho^2} \left(dv- a \sin^2\theta d\psi\right)\ ,\\
	d\tilde\phi&= \frac{(r^2+ a^2)}{\rho^2} \left(d\psi- \frac{a}{a^2+ r^2} dv\right)\ .
\end{aligned}\ee
$d\tilde v$ and $d\tilde\phi$ are the respective analogs of $dt$ and $d\psi$ of the Schwarzschild metric. This correspondence is almost exact with one exception: $d\tilde v$ and $d\tilde\phi$ together with $d\theta$ and $dr$ form an anholonomic basis of one-forms. This means that there are no globally defined coordinates $X$ and $\tilde X$ such that $d\tilde v$ = $dX$ and $d\tilde\phi  = d\tilde X$.
Near the hypersurface $r_0= M+ \sqrt{M^2- a^2}$, we then make the following approximations:
\begin{align}
	\Delta= \frac{\gamma^2 \tilde\epsilon^2}{4}+ \gamma h(v) &M'\ ,\quad r- r_0= \frac{\gamma \tilde\epsilon^2}{4}+ h(v) M'\ , \nonumber\\
	\gamma&= 2 \sqrt{M^2-a^2}\ , \label{bb3}
\end{align}
where $h(v)$ is a function determining the location of Rindler horizon. Then the Kerr--Vaidya metric \eqref{kv} reduces to
\begin{align}
	ds^2&= -\frac{\rho_0^2}{r_0^2+a^2}\left[ \left(\frac{\gamma h}{r_0^2+a^2}- 2\left(1+\frac{2 M}{\gamma}\right)\right)M'\right.\nonumber\\
	&\quad\left.+\frac{\gamma^2 \tilde\epsilon^2}{4(r_0^2+a^2)} \right] d\tilde v^2+ \frac{\rho_0^2 \gamma \tilde\epsilon}{r^2+ a^2} d\tilde v d\tilde\epsilon\\
	&\quad+ \rho_0^2 \left(d\theta^2+  \sin^2\theta d\tilde\phi^2 \right)\ ,\nonumber
\end{align}
where only terms up to order $\tilde\epsilon^2$ and $M'$ have been retained. The function $h(v)$ is now chosen to be
\begin{equation}
	h= \frac{4 r_0 (r_0^2+ a^2)}{\gamma^2}\ ,
\end{equation}
which reduces the above metric to the form
\begin{equation}\label{kv2}
	ds^2= -\frac{\rho_0^2 \gamma^2 \tilde\epsilon^2\ d\tilde v^2}{4(r_0^2+a^2)^2} + \frac{\rho_0^2 \gamma \tilde\epsilon\ d\tilde v d\tilde\epsilon}{r_0^2+ a^2} + \rho_0^2 \left(d\theta^2+  \sin^2\theta d\tilde\phi^2 \right)\ ,
\end{equation}
Finally, we make a further substitution
\begin{equation}
	\rho d\tilde\epsilon= d\epsilon\ ,
\end{equation}
which also implies that
\begin{align}
	d(\rho \tilde\epsilon)
	\approx  d\epsilon\ .
\end{align}
Substituting these relations into \eqref{kv2}, we obtain
\begin{equation}
	ds^2= -\frac{\gamma^2 \epsilon^2\ d\tilde v^2}{4 (r_0^2+a^2)^2} +  \frac{\gamma\epsilon\ d\tilde v d\epsilon}{r_0^2+a^2} + \rho_0^2 (d\theta^2+ \sin^2\theta d\tilde\phi^2)\ . \label{eq62}
\end{equation}

This shows that   the separatrix of a Kerr--Vaidya black hole, determined by \eqref{sep9}, coincides with the Rindler horizon of the Kerr--Vaidya metric. {The temperature associated with the uniformly accelerating Rindler observer can be written as
\begin{equation}
T_U= \frac{\gamma}{4 \pi (r_0^2+a^2)}\ ,
\end{equation}
which is the temperature of the corresponding Kerr--Vaidya spacetime. This procedure for calculating the Hawking temperature, however, is valid only in the quasistatic limit. Ref.~\cite{DS:23} contains arguments detailing how the  temperature associated with the metric~\eqref{eq62} can be extracted.} The Rindler horizon is also the null surface whose parameter rate of area change is constant. To see this, consider the second derivative $d^2 A/d\tau^2$, where $A= 4\pi (r^2+a^2)$ is the surface area along a set of outgoing null rays, using the parametrization $\tau= v$ for null geodesics. The radial null vector for which $d^2 A/dv^2=0$ coincides with that of Eq.~\eqref{sep9} up to order ${\cal O} (M')$. It remains to be seen whether this notion of horizon can play a useful role in generalizing the laws of black hole mechanics to a dynamical setting, where the degeneracy of different horizons is usually absent.

\subsection{Case 2: variable $a$}

We now determine the location of the separatrix for the (ingoing) gKV metric (variable $a$). For this metric also, Eq.~\eqref{eq10} holds. Now, we follow the exact same procedure outlined in the previous section to determine the location of the separatrix
\begin{equation}
	\Delta= 2 (r_0^2+a^2)\frac{M' r_0- a' a}{r_0-M}\ .
\end{equation}
Now, to determine the location of the Rindler horizon, we perform the anholonomic coordinate transformations given in Eq.~\eqref{nc53} to obtain the exact same metric of Eq.~\eqref{nc52} (but with variable $a$). We then make the following approximations:
\begin{equation}
	r- r_0= \frac{\gamma \tilde\epsilon^2}{4}+ h(v) M'+ g(v) a'\ ,
\end{equation}
near the hypersurface $r_0= M+ \sqrt{M^2- a^2}$, where $\gamma$ and $h(v)$ are defined in Eq.~\eqref{bb3}. Here $g(v)$ is an additional function depending on the location of the Rindler horizon. Then the Kerr--Vaidya metric with variable $a$ reduces to
\begin{align}
	ds^2&= -\frac{\rho_0^2}{r_0^2+a^2}\left[ \left(\frac{\gamma h}{r_0^2+a^2}- 2\left(1+\frac{2 M}{\gamma}\right)\right)M'\right.\nonumber\\
	&\quad\left. \left(\frac{\gamma g}{r_0^2+a^2}+ \frac{4 a}{\gamma}\right)a' +\frac{\gamma^2 \tilde\epsilon^2}{4(r_0^2+a^2)} \right] d\tilde v^2\\
	&+ \frac{\rho_0^2 \gamma \tilde\epsilon}{r^2+ a^2} d\tilde v d\tilde\epsilon
	\quad+ \rho_0^2 \left(d\theta^2+  \sin^2\theta d\tilde\phi^2 \right)\ ,\nonumber
\end{align}
where only terms up to order $\tilde\epsilon^2$, $M'$ and $a'$ have been retained. The functions $h(v)$ and $g(v)$ are now chosen to be
\begin{equation}
	h= \frac{4 r_0 (r_0^2+ a^2)}{\gamma^2} ,\quad g= -\frac{4 a (r_0^2+ a^2)}{\gamma^2}\ ,
\end{equation}
which reduces the above metric to the form
\begin{equation}
	ds^2= -\frac{\rho_0^2 \gamma^2 \tilde\epsilon^2\ d\tilde v^2}{4(r_0^2+a^2)^2} + \frac{\rho_0^2 \gamma \tilde\epsilon\ d\tilde v d\tilde\epsilon}{r_0^2+ a^2} + \rho_0^2 \left(d\theta^2+  \sin^2\theta d\tilde\phi^2 \right)\ ,
\end{equation}
Finally, we make a further substitution
\begin{equation}
	\rho d\tilde\epsilon= d\epsilon\ \implies d(\rho \tilde\epsilon)
	\approx  d\epsilon\ ,
\end{equation}
to obtain
\begin{equation}
	ds^2= -\frac{\gamma^2 \epsilon^2\ d\tilde v^2}{4 (r_0^2+a^2)^2} +  \frac{\gamma\epsilon\ d\tilde v d\epsilon}{r_0^2+a^2} + \rho_0^2 (d\theta^2+ \sin^2\theta d\tilde\phi^2)\ .
\end{equation}
Hence, the location of the separatrix of a generalized Kerr--Vaidya black hole also coincides with the Rindler horizon.  {As done above, one can extract the temperature associated with the uniformly accelerating Rindler observer and, consequently, the temperature of the corresponding Kerr-Vaidya spacetime. However, this procedure does not provide an understanding of the thermodynamics of this spacetime beyond the quasi-static limit, and we leave investigation of its utility to future work.}

\section{Discussion} \label{discus}

In this work we considered two classes of Kerr--Vaidya spacetimes as minimal axially symmetric models for the near-horizon region of black holes, based on the fact that (non-rotating) Vaidya metrics provide a self-consistent near-horizon description of physical black holes in the spherically-symmetric case. We made a number of observations of their properties. Noting that Kerr--Vaidya solutions (with constant $a$) which represent evaporating black holes possess timelike apparent horizons, we showed that infalling observers on geodesic trajectories experience no drama during their approach and crossing of the horizon, and may reverse direction if their proper velocity is sufficiently small. Finite NEC violation was shown to occur for infalling test particles which cross the horizon, and in the case of growing black holes, the energy density as measured by Alice will typically be positive. On the other hand, for evaporating black holes it is possible to communicate with the outside world from the apparent horizon and also parts of the trapped region outside the hypersurface $\Delta=0$. This is not so for accreting black holes. In fact, so long as the growth is continuous, the apparent horizon is hidden from outside observers.

On the contrary, we identify previously unappreciated issues with evaporating white hole geometries. We show that for outgoing Kerr--Vaidya metrics with decreasing mass function (counterparts of inadmissible Vaidya solutions) allow for timelike geodesics which reach the singularity while starting  outside of any horizons present. In spherical symmetry this property is not a challenge to the cosmic censorship conjecture, since the relevant solutions may be dismissed as unphysical (due to the impossibility of their realization in finite time according to a distant observer).

For both Kerr--Vaidya black holes and generalized Kerr-Vaidya black holes (those with variable $a$), we also showed that two distinct notions of horizon become equivalent in the near-horizon description. The separatrix of the gKV geometry (which approximates the event horizon when $M'\ll M'^2$) was shown to coincide precisely with the location of the Rindler horizon. This horizon plays a crucial role in the Hawking evaporation of a gKV black hole, in accordance with previous work which used the Rindler form of the Kerr--Vaidya metric (with constant $a$) to determine the associated Hawking temperature. This suggests that the KV and gKV metrics both serve as consistent near-horizon models for capturing back-reaction due to the Hawking process in their respective regimes of applicability.

In this article, we demonstrated the necessity of considering solutions for which angular momentum varies at a different rate from that of the mass. Due to the non-linearity of Einstein's field equations, allowing even a single parameter in the metric to acquire a time dependence significantly increases the complexity of the solutions. The most direct generalization of the Kerr-Vaidya solution, which we termed gKV solutions, involves allowing for a variable angular momentum to mass ratio $a=J/M$. We highlighted the differences and similarities between the properties of variable $a$ solutions with those with constant $a$. We were able to show that gKV solutions are not asymptotically flat unless $a$ approaches a constant value sufficiently rapidly. Thus it will be of interest to investigate the asymptotic properties of metrics that incorporate back-reaction of Hawking radiation on a Kerr background in future work. However, the complexity of variable $a$ solutions makes further analysis difficult. For constant $a$ solutions, there exists a Kodama-like vector field that generates the corresponding Noether current, and its charge coincides with the Brown-York mass in the asymptotically flat region~\cite{AAS:20,Dorau:2024}. The generalization of the Kodama vector to the case of variable angular momentum is not yet known but is left to future investigations.

Like with the spherically symmetric Vaidya metric, both the KV and gKV class of metrics are expected only to provide an accurate description near the horizon, where the flux due to Hawking evaporation is ingoing and of negative energy density. In the far region,  evaporation produces a positive outgoing flux, and a more complete characterization of the geometry necessitates the use of multiple such metrics. We leave the problem of generalizing such multi-Vaidya models  to axially symmetric and cosmological spaces to future investigations.

\acknowledgements

We thank Ioannis Soranidis for useful comments. PKD and SM are supported by an International Macquarie University Research Excellence Scholarship. FS is funded by the ARC Discovery project grant DP210101279. The work of DRT is supported by the ARC Discovery project grant DP210101279.

\appendix

\section{Self-consistent solutions in spherical symmetry} \label{rev-spher}
Here we provide a brief summary  of the self-consistent approach and some key features of physical black holes. Additional details  can be found in Refs. \cite{MMT:22,DSST:24}.  Sec.~\ref{apW} presents the key properties of the white hole solutions.

\subsection{General properties} \label{A:rel}

A general spherically symmetric metric in Schwarzschild coordinates \cite{C:92,vF:15} is given by
\be
ds^2 = -e^{2h(t,r)}f(t,r)dt^2+f(t,r)^{-1}dr^2+r^2d\Omega_2\ . \label{m:tr}
\ee
In terms of the advanced null coordinate $v$ the metric is
\be
ds^2=-e^{2h_+(v,r)}f_+(v,r)dv^2+2e^{h_+(v,r)}dvdr+r^2d\Omega_2\ , \label{m:vr}
\ee
and using the retarded null coordinate $u$ it is written as
\be
ds^2=-e^{2h_-(u,r)}f_-(u,r)dv^2-2e^{h_-(u,r)}dudr+r^2d\Omega_2\ . \label{m:ur}
\ee
The function $f$ is coordinate-independent, i.e. $f(t,r)\equiv f_+\big(v(t,r),r\big)$ and in what follows we omit the subscript. It is conveniently represented via the Misner--Sharp--Hernandez (MSH) mass $M\equiv C/2$ as
\be
\pad_\mu r \pad^\mu r\equiv f\eqdef 1-\frac{C(t,r)}{r}=1-\frac{C_+(v,r)}{r}\ ,
\ee
where the coordinate $r$ is the areal radius \cite{vF:15}. 		
The functions $h$ and $h_\pm$ play the role of integrating factors in coordinate transformations, such as Eq.~\eqref{urtr} below.

Analysis of the Einstein equations and the evaluation of curvature invariants are  conveniently performed using the   effective EMT components $\tau_a$, (where $a={}_t, {}^r, {}_t{}^r $)   \cite{MMT:22}
\begin{align}
	\tau{_t} \defeq e^{-2h} {T}_{tt}\ , \qquad {\tau}{^r} \defeq T^{rr}\ , \qquad  \tau {_t^r} \defeq e^{-h}  {T}{_t^r}\ . \label{eq:mtgEMTdecomp}
\end{align}
The Einstein equations for the components $G_{tt}$, ${G_t}^r$, and ${G}^{rr}$ are then, respectively
\begin{align}
	\partial_r C &= 8 \pi r^2  {\tau}{_t} / f\ , \label{eq:Gtt} \\
	\partial_t C &= 8 \pi r^2 e^h  {\tau_t}^r\ , \label{eq:Gtr} \\
	\partial_r h &= 4 \pi r \left(  {\tau}{_t} +  {\tau}{^r} \right) / f^2\ . \label{eq:Grr}
\end{align}
The apparent horizon is the
boundary of this trapped region and it corresponds to the domain $f<0$ in Eq.~\eqref{m:vr} \cite{HE:73,vF:15}.

In an asymptotically flat spacetime the Schwarzschild radius $r_\sg$ is the largest root of $f(t,r)=0$ (see Ref.~\cite{MMT:22} and the references therein for the detailed summary of the definitions and their consequences).
Invariance of the MSH mass implies that
\be\label{msh}
r_\sg(t)=C(t,r_\sg)=r_+\big(v(t,r_\sg(t))\big)\ ,
\ee
where $r_+(v)$ is the largest root of $f_+(v,r)=0$. It represents the location of the outer component of the apparent horizon of a black hole or the anti-trapping horizon of a white hole. Unlike the globally defined event horizon, the notion of the apparent horizon is foliation-dependent. However, it is invariantly defined in all foliations that respect spherical symmetry \cite{FEFHM:17}.


Our regularity requirement is the weakest form of the cosmic  censorship conjecture \cite{FN:98,vF:15}: all polynomial invariants that are based on the contractions of the Riemann tensor \cite{HE:73,SKMHH:03} are finite up to and on the apparent horizon. Constructing  finite invariants from the divergent quantities that describe a real-valued solution allows one to describe properties of the near-horizon geometry. It is sufficient to ensure that only two of them, $R$ and $R_{\mu\nu}R^{\mu\nu}$, are finite \cite{T:19}. Moreover, we focus on the quantities,
\begin{align}
& \mathrm{T}\defeq( {\tau}{^r} -  {\tau}{_t}) / f, \\
&  \mathfrak{T} \defeq \big( ( {\tau}{^r})^2 + ( {\tau}{_t})^2 - 2 ( {\tau}{_t^r})^2 \big) / f^2\ ,
\end{align}
which are the potentially divergent parts of the full expressions for the curvature scalars
\be
T^\mu_{~\mu}=\mathrm{T} +2T^\theta_{~\theta}, \qquad  T^{\mu\nu}T_{\mu\nu}=\mathfrak{T}+ 2\big(T^\theta_{~\theta}\big)^2.
\ee
The contributions from $T^\theta_{~\theta}\equiv T^\phi_{~\phi}$ can initially be disregarded, as one can verify that they do not introduce further divergences \cite{MMT:22,T:19}. Because the metric in Schwarzschild coordinates is singular at the apparent horizon, working instead with $(u,r)$ and $(v,r)$ coordinates allows one to identify the resulting solutions. Note that in $1+1$ dimensions the condition of regularity at the horizon is known to be equivalent to its formation in finite time, however it is not known whether this is the case in $3+1$ dimensions \cite{BL:08}.

The assumptions of finite formation time and regularity restrict the scaling of the effective EMT components near the Schwarzschild radius, such that $\tau_a\propto f^k$, with $k=0,1$. Solutions with $k=0$ describe a PBH after  formation (and before a possible disappearance of the trapped region).  Vaidya metrics, and dynamical regular black hole solutions belong to this class \cite{MS:23}, while the Reissner-Nordstr\"{o}m solution or static RBH solutions correspond to $k=1$~\cite{sm:23}.  In the following we work with $k=0$ solutions.

The EMT components of the  $k=0$ solutions satisfy
\be
\lim_{r\to r_\sg} \tau_t=\lim_{r\to r_\sg} \tau^r=-\Upsilon^2(t)\ , \quad \lim_{r\to r_\sg} \tau^r_t=\pm \Upsilon^2(t)\ ,
\ee
for some function $\Upsilon(t)>0$.
The leading terms of the metric functions in a near-horizon expansion are given in terms of $x\defeq r-r_\sg(t)$ as
\begin{align}
	C &= r_\sg - 4 \sqrt{\pi} r_\sg^{3/2} \Upsilon \sqrt{x} + \mathcal{O}(x)\ , \label{k0C} \\
	h &= - \frac{1}{2}\ln{\frac{x}{\xi}} + \mathcal{O}(\sqrt{x})\ . \label{k0h}
\end{align}
The function $\Upsilon(t)$ determines the energy density, pressure and flux at the apparent horizon, and  $\xi(t)$ is determined by choice of the time variable. 

The Einstein equation~\eqref{eq:Gtr} serves as a consistency condition and establishes the relation between the rate of change of the MSH mass and the leading terms of the metric functions,
\begin{align}
	r'_\sg/\sqrt{\xi} =   \pm 4\sqrt{\pi r_\sg} \, \Upsilon\ . \label{eq:k0rp}
\end{align}
The   EMT near the Schwarzschild sphere is
\begin{align}
	\ T^a_{~b} = \begin{pmatrix}
		\Upsilon^2/f &   \pm e^{-h}\Upsilon^2/f^2 \vspace{1mm} & 0 & 0\\
		\mp e^h  \Upsilon^2 & -\Upsilon^2/f & 0 &0 \\
 0 & 0 & p_\| &0 \\
 0 & 0 & 0 & p_\|
	\end{pmatrix}\ ,
		\label{tneg}
\end{align}
where the upper sign corresponds to the (evaporating) PBH and the tangential pressure $p_\|$ is finite at $r=r_\sg$. Both signs lead to the violation of the NEC. In particular this means that out   of the four possible types of Vaidya metrics, $h_\pm=0$, $f=1-r_+(v)/r$, or $f=1-r_-(u)/r$, only the solutions with $r_+'<0$ and $r_-'>0$  are allowed \cite{MMT:22}. The retreating apparent horizon and the advancing anti-trapping horizon are timelike surfaces \cite{BMMT:19}.

Consider the energy density $\varrho=T_{\mu\nu}u^\mu u^\nu$, the pressure $p=T_{\mu\nu}n^\mu n^\nu=T^r_{~r}$, and the
flux $\varphi= T_{\mu\nu}u^{\mu}n^\nu$  that are perceived by various observers with four-velocities $u^\mu$   and  outward-pointing radial spacelike vector $n^\mu$. For an observer at constant $r$ all of these quantities, $\varrho=-T^t_{~t}$, $p=T^r_{~r}$, diverge in the limit $r\to r_\sg$. 



\subsection{White holes}\label{apW}

A white hole is an anti-trapped region --- a  domain where both
ingoing and outgoing future-directed null geodesics emanating
from a spacelike two-dimensional surface with
spherical topology have positive expansion. The anti-trapping horizon is the
boundary of this region.

 Admissible white hole solutions in the self-consistent approach  correspond to $r_\sg'>0$ and are conveniently described using $(u,r)$ coordinates \cite{BMMT:19,MMT:22D}. The retarded null coordinate $u$,
 \be
dt=e^{-h}(e^{h_-}du+ f^{-1}dr)\ , \label{urtr}
\ee
is regular across the expanding anti-trapping horizon.
 In analogy with PBHs, we refer to white holes that form in finite time according to Bob as physical white holes (PWHs). The name, however, does not imply that these objects are necessarily found in nature.

A useful relationship between the EMT components $T_{\mu\nu}$ in $(t,r)$ coordinates and $\Theta_{\mu\nu}$ in $(u,r)$ coordinates is given by
\begin{align}
	&	 {\theta}{_u} \defeq e^{-2h_-}  {\Theta}_{uu} =  {\tau}{_t} , \label{eq:thev} \\
	&	 {\theta}_{ur} \defeq  e^{-h_-}  {\Theta}_{ur} = \left(  {\tau}{_t^r} +  {\tau}{_t} \right) / f , \label{eq:thevr}\\
	&	 {\theta}{_r} \defeq  {\Theta}_{rr} = \left(  {\tau}{^r} +  {\tau}{_t} + 2  {\tau}{_t^r} \right) / f^2 ,   \label{eq:ther}
\end{align}
where similarly to the Schwarzschild coordinate setting we introduced the effective EMT components $\theta_a$, $a=u,r,ur$.  The Einstein equations can then be written as
\begin{align}
	&  \pad_u C_-  = 8 \pi r^2  \Theta^r_{\;u}= -8 \pi r^2 e^{h_-}( {\theta}{_u}- f  {\theta}_{ur}) , \label{eq:Gvv}\\
	& \pad_r C_- =8\pi r^2 \Theta^u_{\;u}=  - 8 \pi r^2  {\theta}_{ur}  , \\
	& \pad_r h_- =  -4 \pi r e^{h_-}\Theta^u_{\;r}= 4 \pi r  {\theta}{_r} . \label{vGrr}
\end{align}

Tangent vectors to the congruences of  outgoing and ingoing radial null geodesics   are given in $(u,r)$ coordinates by
\begin{align}
	l_+^\mu=(0, e^{-h_-},0,0), \qquad l_-^\mu=(1,-\half e^{h_-}f,0,0), \label{null-v}
\end{align}
respectively.   The vectors are normalized to satisfy $l_+\cdot l_-=-1$. Their expansions  are
\begin{align}
	\vartheta_{+} =   \frac{2e^{-h_-}}{r}, \qquad \vartheta_{-}=-\frac{e^{h_-}f}{r},
\end{align}
respectively. Hence the (outer) anti-trapping  horizon is located at the Schwarzschild radius $r_\sg$, justifying the definition of the white hole mass as $2M(u)=r_-(u)$.

Using the Einstein equations and the relationships between components of the EMT in two coordinates systems  one can show that
\begin{align}
& 2M(u)\defeq C_-(u,r)=r_-(u)+w_1(u) y +\ldots\ ,  \\
& h_-(u,r)=\chi_1(u) y +\ldots\ ,          \label{lfu}
\end{align}
where   $r_-(u)$ is the radial coordinate of the apparent horizon, we define $y\defeq r-r_-(u)$, and $C_-(u,r_-)\equiv r_-$.   The Vaidya geometry with $C'_-(u)>0$ and $h_-\equiv 0$ is the simplest example of such a white hole.

We focus now on Vaidya solutions as they capture all the near-horizon features of solutions of the same type.
The four-velocity of a radially moving observer Alice is
   \be
 u^\mu_\mathrm{A}=\big(\dot u,  \dot r,  0,0\big),
 \ee
and the spatial outward-pointing  unit vector normal to it is
\be
n^\alpha=(-\dot u, f\dot u+\dot r,0,0,).
\ee
If $\dot r^2\gg f$, then  expanded in terms of $z=r-r_-$ gives
\be
\dot u=-2\dot r\frac{ r_-}{z}+\cO\big(z^0\big), \label{uIn}
\ee
for ingoing observers if $z\ll r_-$, and
\be
  \dot u\approx -\frac{2\dot r}{1-\chi_+\alpha}=-\frac{\dot r}{\alpha}+\cO(\alpha) \label{uInL}
\ee
for the linear evaporation law with $\alpha<1/8$. On the other hand,
\be
\dot u=\frac{1}{2\dot r^2}+\cO(z),  \label{uOut}
\ee
for outgoing observers.

 The only nonzero EMT component is $T_{uu}=-r'_-/(8\pi r^2)$, and thus Alice's locally measured energy density, pressure and flux are
\be
\varrho_\rA=-\varphi_\rA=p_\rA=-\frac{r'_-(u)\dot u^2}{8 \pi r^2 }\ .	
\ee
Then using Eq.~\eqref{uOut} we obtain Eq.~\eqref{rhoOut}, while for the ingoing observer with $\dot r^2\gg f$ Eq.~\eqref{uIn} gives
\be
\varrho_\rA=-\varphi_\rA=p_\rA=-\frac{r'_-(u)\dot r^2}{2 \pi z^2 }+\cO\big(z^{-1}\big) .	\label{emTin}
\ee

\begin{figure*}[!htbp]
  \centering
  \begin{tabular}{@{\hspace*{0.025\linewidth}}p{0.45\linewidth}@{\hspace*{0.05\linewidth}}p{0.45\linewidth}@{}}
  	\centering
  	\subfigimg[scale=0.40]{(a)}{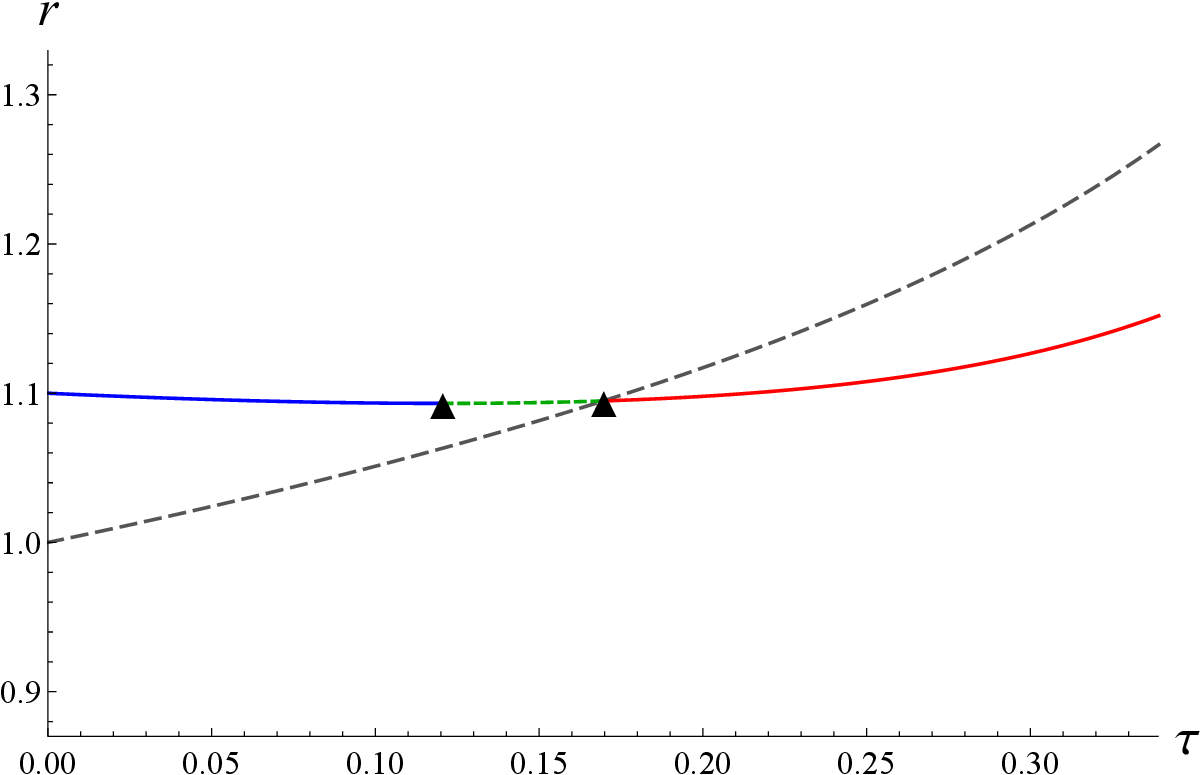} &
  	\subfigimg[scale=0.40]{(b)}{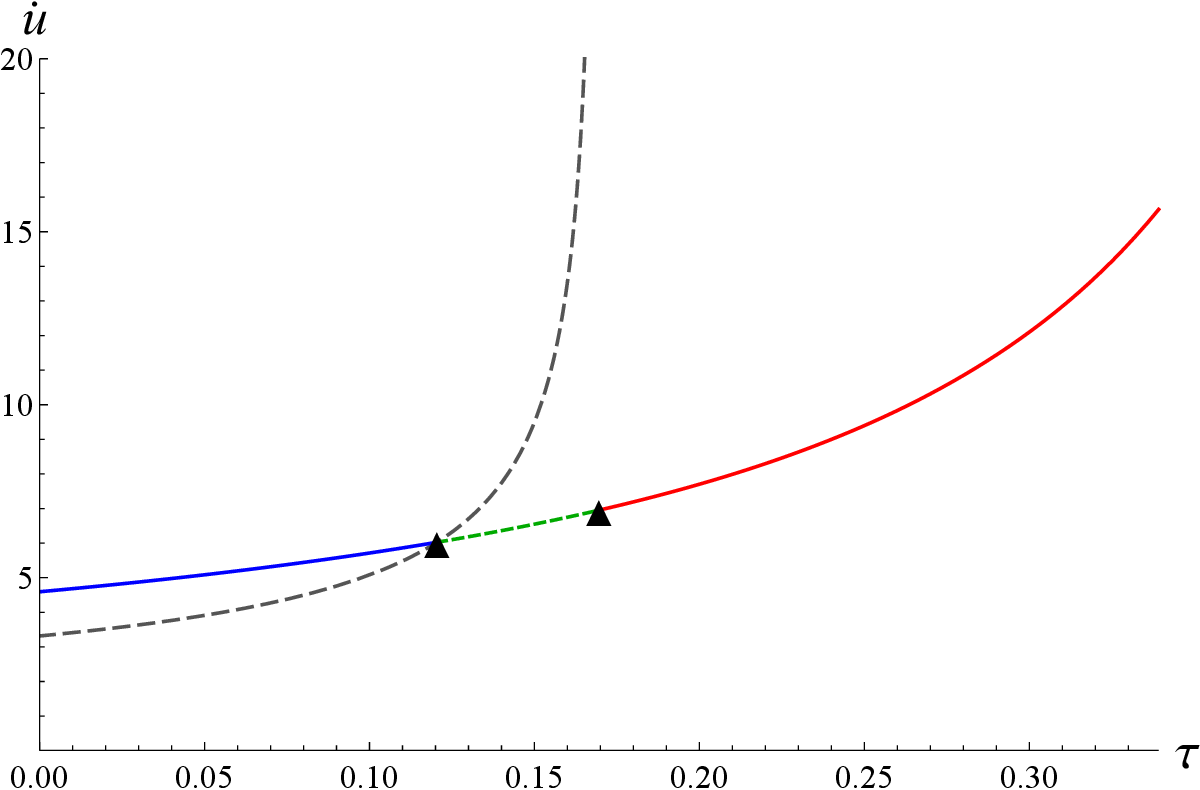}
  \end{tabular}
  	\caption{Entry of a massive test particle into the Vaidya white hole. Both figures are based on the linear accretion law $r_-(u)=r_-(0)+\alpha u$ with $r_-(0)=1$ and $\alpha=0.1$ The initial conditions are $u(0)=0$, $r(0)=1.1$ and $\dot r(0)= -0.1$. (a) Trajectory from the initial moment until the reversal and the capture. The areal radius $r(\tau)$ and the (outer) apparent horizon $r_{-}\big(u(\tau)\big)$ (gray dashed line) are shown as   functions of the proper time $\tau$.  (b) Segments of the trajectory showing the ``reversal'' (the geodesic switches from being outgoing to ingoing) and capture. The time derivative $\dot u$ is shown as a solid line, and the limiting value $\sqrt{f\big(u(\tau),r(\tau)\big)}$ as a dashed gray line. Here, blue line represent infalling trajectories, dotted green line represent outgoing trajectories outside the horizon and red line represent trajectories after the capture. }
  	\label{velR}
\end{figure*}

The anti-trapped region of a classical (``eternal") white hole is inaccessible to external observers. However, as the anti-trapping horizon of a PWH at $r_\sg\equiv r_-$ is timelike \cite{BMMT:19}, test particles can cross it.

The motion of timelike test particles is more intricate. Outside of the anti-trapped region, both ingoing and outgoing radial trajectories with four-velocity $u_\rA=(\dot u, \dot r, 0, 0)$ satisfy the relation
\be
\dot u=\frac{-\dot r +\sqrt{\dot r^2+f}}{f}\ , \label{inf}
\ee
where $f=1-r_-\big(u(\tau)\big)/r(\tau)$. This follows from the timelike condition $u_\rA^\mu u_{\rA\,\mu}=-1$ and the future-directedness of the coordinate $u$, namely $\dot u>0$. The geodesic equations for radial timelike geodesics are
\begin{align}
    & \ddot r=-\frac{r_-}{2r^2}+\frac{r'_-}{2r}\dot u^2\ , \label{eqR}\\
   & \ddot u= \frac{r_-}{2r^2}\dot u^2\ ,    \label{eqU}
\end{align}
where the first term on the right-hand side of Eq.~\eqref{eqR} is absent for null geodesics.

Inside the region $f<0$ the areal radius can only grow and thus the radial velocity component is positive ($\dot r>0$) which is consistent with having $\ddot r>0$ near $r_-$. As a result,  there are two a priori possible entry scenarios: either $\dot r(\tau)$ becomes zero upon entry at $r(\tau)=r_-\big(u(\tau)\big)$, or at some $r=\tilde{r}>r_-$ the infall stops, the particle becomes outgoing, and is then overtaken by the expanding horizon. We will now show that only the latter option is possible. Assume, to the contrary, that the radial velocity  goes to zero at the horizon crossing. Then
\be
\lim_{r(\tau)\to r_-(u(\tau))} \frac{\dot r^2}{f}=A\ , \label{rhoOut}
\ee
where $A$ depends on the initial conditions and a priori can be either finite  or infinite. In both cases   $\dot u$ and $\ddot u$ diverge. However, the divergences obtained from $\ddot u$ as determined by the geodesic equation \eqref{eqU}, and as calculated directly from the proper time derivative of $\dot u$, are not consistent with each other. For example, if $A=0$, then according to Eq.~\eqref{eqU} $\ddot u$ should diverge  as $f^{-1}$, while  Eq.~\eqref{inf} implies an $f^{-2}$ divergence. Hence the radial velocity cannot go to zero at the horizon, and the only valid entry scenario involves the particle turning around and being overtaken by the expanding horizon.

We illustrate this using  a simple model of a linearly expanding Vaidya PWH. Fig.~\ref{velR} shows the  trajectory of a massive, initially ingoing test particle.   Eq.~\eqref{inf}  implies that for ingoing particles $\dot u\geqslant 1/\sqrt{f}$.  This value is reached when $\dot r= 0$, at which point the particle reverses direction and starts moving radially outward, where it may be overtaken by the expanding apparent horizon. The associated energy density, flux and pressure in the proper frame of the test particle when it enters the white hole are  finite (see Appendix~\ref{A:rel}),
\begin{align}
	\varrho_\rA=-\varphi_\rA=p_\rA=-\frac{r'_-(u)}{8 \pi r^2\dot{r}^2}\ .
\end{align}

White hole and black hole solutions can be conveniently described as time reverses of each other. For Vaidya metrics this is accomplished by taking $v\to -u$. The complicated entry scenario described above has a counterpart in the exit of test particles from an evaporating Vaidya black hole. There, outgoing geodesics are reversed and become ingoing, and subsequently  may be overtaken by the contracting apparent horizon \cite{DST:22}.

However, some of the usual interpretations of the role of the Vaidya metric in modelling with $r'(v)<0$ evaporating PBHs  and do not directly translate to the white holes with $r'(u)>0$. In both cases the event horizon is hidden by the apparent/anti-trapping horizon, but in the black hole case the ingoing Vaidya metric with $r'(v)<0$ is consistent with the requirement that one has a future horizon and a decreasing mass due to Hawking evaporation. Such an approximation is usually taken to be valid near the horizon ($r\lesssim   r_\sg$) where the NEC violation is interpreted as a flux of negative energy (as defined by a timelike Killing vector in the asymptotic region) across the horizon, which serves to reduce the black hole mass \cite{BMPS:95,FN:98}.

The corresponding white hole geometry contains a near-horizon region described by an outgoing Vaidya metric with $r_-'(u)>0$, as the evaporating case does not correspond to the finite formation time according to distant Bob.  However, the physical interpretation of the NEC violating null fluid in the black hole case does not readily carry over to the white hole geometry. Since $r_-'(u)>0$, the outgoing flux must be of negative energy modes in order for the mass to increase. However, contrary to the black hole case where such modes tunnel from a classically forbidden region (outside the horizon) to one where their momenta are timelike (inside the horizon) \cite{FN:98,BMPS:95,M:15}, the modes here would be of negative energy outside the horizon.

 As a result, the interpretation of Hawking radiation as a tunnelling process across the horizon is incompatible with the consistency conditions we impose, when applied to the  semiclassical white hole geometry. This is not entirely surprising given the numerous pathologies that arise when describing the Hawking process in a white hole background. In the (eternal) black hole background, the late-time thermal density matrix constituting Hawking radiation is insensitive to ambiguities in defining a notion of positive frequency for horizon states. For the white hole geometry the state of the field observed at future null infinity $\mathcal{I}^+$ is quite arbitrary, depending strongly on the choice of state in the asymptotic past $\mathcal{I}^-$ and on the past horizon $\mathcal{H}_-$. A simple time-reversal of the Unruh state defined in the evaporating black hole case produces a state with an unnaturally high degree of correlation between $\mathcal{I}^-$ and $\mathcal{H}_-$ and no flux at $\mathcal{I}^+$, while more natural choices produce divergent outgoing fluxes which invalidate the semiclassical approximation \cite{WR:80}.

\section{Kerr--Vaidya metrics}

\subsection{Expansions of the outgoing and ingoing KV metric}           \label{aC}
Here we  focus on the black hole solutions using the $(v,r)$ coordinates. Expansions of the outgoing and the ingoing geodesics that are tangent to the normals to the hypersurface $\Delta=0$
\begin{align}
l^+_\mu=& \frac{\rho^2}{2\Sigma^2}\big(-\Delta, r^2+a^2+\Sigma,0,0\big), \\
l^-_\mu=& \big(-1, (r^2+a^2-\Sigma)/\Delta,0,0\big),
\end{align}
have the expansions
\begin{align}
\vartheta_+=& \frac{1}{2\Sigma^3} \bigg(\Delta(2 r\rho^2+ a^2\sin^2\theta (M+r)) \nonumber\\
&+ M' r a^2\sin^2\theta (r^2+ a^2+\Sigma)\bigg)\ ,\\
\vartheta_-=& -\frac{1}{\Sigma\rho^2} \bigg(2 r\rho^2+ a^2\sin^2\theta (M+r) \nonumber\\
&+ \frac{r^2+a^2-\Sigma}{\Delta} M' r a^2\sin^2\theta\bigg)\ ,
\end{align}
 respectively \cite{ST:15}. This identifies the Kerr--Vaidya solutions as black holes and  indicate that the trapped region extends beyond $\Delta=0$ in the evaporating case. Its boundary as described in Eq.~\eqref{ah-kv} can be determined
using the expansions of the pair
of outward- and inward- pointing future-directed null vector orthogonal to the $u=\text{constant}$ hypersurface and the two-surface $r_0+M'\varsigma(v,\theta)$ that (before the rescaling) are given by \cite{DT:20}
\begin{equation}
    \ell^\pm_\mu \propto (-1,\ell_\pm , - \ell_\pm M'\pad_\theta\varsigma,0)\ .
\end{equation}
The null condition $l^\pm\cdot l^\pm=0$ gives
\be
    l^\pm =
    \frac{1}{\Delta+(M'\pad_\theta \varsigma)^2}\Big(r^2+a^2\pm\sqrt{\Sigma^2-(aM'\pad_\theta\varsigma)^2\sin^2\theta}\,\Big)\ .
\ee

\subsection{Energy-momentum tensor  }    \label{aB}

\subsubsection{Kerr--Vaidya metrics}
The non-zero components of the EMT for the Kerr--Vaidya metric \eqref{kv} in advanced coordinates are
 \begin{align}
&  T_{v v}=\frac{ r^2(a^2+ r^2)- a^4 \cos^2\theta\sin^2\theta}{4\pi\rho^6}M_v
-\frac{a^2 r \sin^2\theta}{8\pi\rho^4}M_{vv}\ ,\\
 & T_{v \theta}=-\frac{ a^2 r \sin\theta \cos\theta}{4\pi \rho^4}M_v,\\
& T_{v \psi} = - a \sin^2\theta T_{v v}-a \sin^2\theta\frac{r^2-a^2\cos^2\theta}{8\pi\rho^4}M_v\ , \\
& T_{\theta \psi}=\frac{ a^3 r \sin^3\theta \cos\theta}{4\pi \rho^4}M_v\ ,\\
& T_{\psi \psi}= a^2 \sin^4\theta T_{v v}+ a^2 \sin^4\theta\frac{r^2-a^2\cos^2\theta}{4\pi\rho^4}M_v\ . \end{align}
Likewise, the non-zero components of the EMT for the Kerr--Vaidya metric \eqref{ku} in retarded coordinates are
\begin{align}
& T_{u u}=-\frac{r^2( a^2 +r^2)- a^4 \cos^2\sin^2\theta}{4\pi\rho^6}M_u
-\frac{a^2 r \sin^2\theta}{8\pi\rho^4}M_{u u}\ , \label{c4}\\
& T_{u \theta}=-\frac{2 a^2 r \sin\theta \cos\theta}{8\pi \rho^4}M_u,\\
&  T_{u \psi}=-  a\sin^2\theta T_{u u}+a\sin^2\theta \frac{r^2-a^2\cos^2\theta}{8 \pi\rho^4}M_u\ ,\\
& T_{\theta \psi}=\frac{2 a^3 r \sin^3\theta \cos\theta}{8\pi \rho^4}M_u\ ,\\
&  T_{\psi \psi}= a^2 \sin^4\theta T_{u u}-a^2 \sin^4\theta\frac{(r^2-a^2\cos^2\theta)}{4\pi\rho^4}M_u\ . \end{align}

In advanced coordinates the decomposition \eqref{MT-emt} of the EMT is obtained using the vectors
\be
k_\mu=(1,\; 0,\; 0,\; -a\sin^2\theta), \
\ee
and
\be
q_\mu=\bigg(0,\; 0,\; T_{v \theta},\; -a\sin^2\theta\frac{r^2-a^2\cos^2\theta}{8\pi\rho^4}M_v\bigg).
\ee
The orthonormal tetrad for which $k^\mu=e_{\hat 1}^\mu+e_{\hat 0}^\mu$ is given by
\begin{align}
&e_{\hat 0}^\mu= (-1,rM/\rho^2,0,0)\ ,    \label{b136} \\
&e_{\hat 1}^\mu= (1,1-rM/\rho^2,0,0)\ ,  \\
&e_{\hat 2}^\mu=(0,0,1/\rho,0)\ , \\
&e_{\hat 3}^\mu=\frac{1}{\rho}\big(a\sin\theta,a\sin\theta,0,\csc\theta\big)\ . \label{b139}
\end{align}
Hence the EMT is given by Eq.~\eqref{IVP} with
\be
\nu=T_{vv}\ ,\qquad q^\mu=q^{\hat 2} e_{\hat 2}^\mu+q^{\hat 3} e_{\hat 3}^\mu\ ,
\ee
where
\begin{align} q^{\hat 2}=& -\frac{a^2 r M_v}{8\pi \rho^5} \sin2\theta\ , \\
q^{\hat 3}=&{ -\frac{a(r^2 - a^2\cos^2\theta) M_v}{8\pi \rho^5}\sin\theta\ .}
\end{align}

\subsubsection{Generalised Kerr--Vaidya metrics} \label{aB-var}

Although the general form of the EMT is not informative, we can take the limit as $r\to\infty$ to study the asymptotic behaviour of the EMT of such a spacetime. For the ingoing gKV metric the non-zero components are
\begin{align}
T_{v\phi}&=-\frac{1}{2} \sin ^2\theta a''(v)\ ,\\
T_{\theta\theta}&=\frac{1}{16} \left(16 a \sin ^2\theta a''(v)+12 \sin ^2\theta a'(v)^2\right)\ ,\\
T_{\phi\phi}&=\frac{1}{8} \left(8 a \sin ^4\theta a''(v)+2 \sin ^4\theta a'(v)^2\right)\ .
\end{align}
Let $T^{\ha\hat{b}}$ be the EMT in an orthonormal frame defined by Eqs.~\eqref{b136}-\eqref{b139} with $a=a(v)$. The Lorentz-invariant eigenvalues are then the roots of the equation
\be
\det (T^{\ha\hat{b}}-\lambda \eta^{\ha\hat{b}})=0, \qquad \eta^{\ha\hat{b}}=\mathrm{diag}(-1,1,1,1)\ .
\ee
For the above EMT we have four distinct eigenvalues, two of which are real and two of which are complex. As the general forms of the eigenvalues are lengthy, we show explicit expressions for two of the real eigenvalues around the equatorial plane. They are given by
\begin{align}\label{b31}
\lambda^{(1)}=& \frac{\mathfrak{l}^{1/3}}{3^{2/3} r^8}-\frac{r^3 a'(v) \left( (4 M+r) a'(v) +2 a M'(v)\right)}{(3 \chi)^{1/3}}\ ,\\
\lambda^{(2)}=& -\frac{a a'(v)}{r^3} ,
\end{align}
where subleading terms of order ${\cal O}(M'^2,a'^2)$ have been discarded and we have defined
\begin{align}
\mathfrak{l}&= 9 r^{21} a'(v)^2 M'(v)+9 r^{19} a a'(v)^2 \left(2 M a'(v)+a M'(v)\right) \nonumber\\
&\quad+\bigg[3 r^{38} a'(v)^3 \bigg(27 a'(v) \left(2 a M a'(v)+\left(a^2+r^2\right) M'(v)\right)^2 \nonumber \\
&\quad+r \left(a'(v) (4 M+r)+2 a M'(v)\right)^3\bigg)\bigg]^{1/2}\ .
\end{align}
In the most general scenario, there exist four distinct eigenvalues, with two of them being complex. The remaining two eigenvalues are real and correspond to spacelike eigenvectors. As a result, we have no real timelike or null eigenvectors. This implies that the EMT is of type IV. The real spacelike eigenvector is $e_{\hat 2}^\mu$ corresponding to the eigenvalue $\lambda^{(2)}$. Similarly, the spacelike eigenvector corresponding to the eigenvalue $\lambda^{(1)}$ is
\be
\bigg(1,1-\dfrac{\lambda^{(1)} r^4 a'(v)}{\zeta}, 0,\dfrac{-\lambda^{(1)2} r^6}{\zeta}\bigg)\ ,
\ee
where
\begin{align}
\zeta&=a'(v) \left(\lambda^{(1)} r^3 (2 M+r)-2 \left(a^2+r^2\right) M'(v)\right)\nonumber\\
&\quad-4 a M a'(v)^2+\lambda^{(1)} r^3 a M'(v)\ .
\end{align}
 {As $a'(v)\to 0$, we can see from Eq. \eqref{b31} that both real eigenvalues of EMT goes to zero. Moreover, explicit calculations show that remaining two complex eigenvalues also go to zero. The eigenvectors in this case thus reduces to
\begin{equation}
    (1,1,0,0), \quad \big(0,0,1,-q^{\hat 2}/q^{\hat 3}\big)\ ,
\end{equation}
which are null and spacelike, respectively. Thus, the Kerr--Vaidya metric with $a=\mathrm{const.}$ is of type III.}

\subsection{Asymptotic flatness conditions for gKV metrics}
The gKV metrics satisfy a naive notion of asymptotic flatness: they approach the Minkowski metric with the correction scaling as $\cO(r^{-1})$ for large $r$ \cite{HE:73,W:84}. However, as their EMT does not vanish as $r\to\infty$ (Sec.~\ref{aB-var}), the geometry is not asymptotically empty even in the weakest possible sense \cite{HE:73}.

The modern view of asymptotic flatness \cite{W:84,F:04,ANK:12} centers on conformal compactification, where the metric is rescaled by a conformal factor that approaches zero at infinity. This rescaling makes infinity finite within the auxiliary metric, allowing the points at infinity in the physical spacetime to be effectively added as a boundary. A spacetime is considered asymptotically flat if this boundary can be included in an appropriately regular manner, with the auxiliary space exhibiting properties similar to Minkowski spacetime at its conformal boundary.
We examine evaporating gKB black holes which are assumed to approach a Kerr metric as $v\to -\infty$ to avoid having infinite mass and angular momentum. Hence our focus is on the mapping in the vicinity of future null infinity $\mathcal{I}^+$,  and thus on future asymptotic flatness.

The asymptotic flatness of the gKV solutions is most easily determined by studying the  Newman–Penrose scalars \cite{C:92,SKMHH:03}. Their asymptotic behaviour is given below. The five complex scalars that represent the Weyl tensor scale as
\begin{align}
    & \Psi_0 = 0\ ,\\
    & \Psi_1 = \frac{i a^2 a' \cos \theta \sin2\theta}{4 r^4}-\frac{a a' \sin2\theta}{8r^3} + \mathcal{O}\left(r^{-5}\right)\ ,\\
    & \Psi_2 = \frac{a \sin^2\theta a''}{12r^2}-\frac{ia'\cos\theta}{2r^2}   +\mathcal{O}\left(r^{-3}\right)\ ,\\
    & \Psi_3 = \frac{i\sin\theta a''}{4r} +  \mathcal{O}\left(r^{-3}\right)\ ,\\
    & \Psi_4 = -\frac{\sin^2\theta(2a'^2 + a a'')}{2 r^2}+\mathcal{O}\left(r^{-3}\right)\ .
\end{align}
Despite presence of some non-decaying components of the EMT \cite{CGK:90} that are explicitly given in Section \ref{aB-var}, all go to zero. The four real scalars
\begin{align}
     \Phi_{00}&=0\ ,\\
     \Phi_{11} &= \frac{3 a a' + a a' cos2\theta}{8r^3}-\frac{\sin^2\theta a'^2}{16 r^2}  + \mathcal{O}\left(r^{-4}\right)\ ,\\
     \Phi_{22} &= \frac{1}{4r^2}\left(6a'^2 + 2 \cos2\theta a'^2 - 8M'\right.\notag\\
    &\left.\qquad + 5a a'' + 3 a a'' \cos2\theta\right) + \mathcal{O}\left(r^{-3}\right)\ ,\\
    \Lambda &=\dfrac{R}{24}= \frac{(4a'^2 +3a a'') \sin^2\theta}{48r^2} +\mathcal{O}\left(r^{-3}\right)\ ,
\end{align}
and three complex scalars
\begin{align}
    & \Phi_{01} = -\frac{i a'\sin\theta}{4 r^2} + \frac{a a' \cos\theta \sin\theta}{2r^3}  + \mathcal{O}\left(r^{-4}\right) ,\\
     & \Phi_{02} = -\frac{a'^2 \sin^2\theta+2aa'' \sin^2\theta}{8r^2}   + \mathcal{O}\left(r^{-3}\right) ,\\
    & \Phi_{12} = -\frac{i a''\sin\theta}{4 r} - \frac{i a' \sin\theta}{4 r^2}  + \mathcal{O}\left(r^{-3}\right) ,
\end{align}
representing the Ricci tensor all approach zero for fixed $v$ (or $u$) as $r\to\infty$. However, future asymptotic flatness requires that these quantities approach zero sufficiently rapidly as $r\rightarrow\infty$. The relevant conditions \cite{B:60,S:62,F:04,ANK:12} on the leading terms are
\begin{align}
		 \Psi_0 & =\psi_0^0\,r^{-5} +\cO\big(r^{-6}\big)\ ,\\
		 \Psi_1 & = \psi_1^0\,r^{-4} +\cO\big(r^{-5}\big)\ ,\\
		 \Psi_2 & = \psi_2^0\,r^{-3} +\cO\big(r^{-4}\big)\ ,\\
		 \Psi_3 & = \psi_3^0\,r^{-2} +\cO\big(r^{-3}\big)\ ,\\
		 \Psi_4 & = \psi_4^0\,r^{-1} +\cO\big(r^{-2}\big)\ ,
\end{align}
where the functions $\psi_k^0$ ($k\in\{0,...,4\})$ depend only on the retarded coordinate $u$ and the angular coordinates. The radial coordinate $r$ can be used to parameterize null geodesics.
Thus, it appears that while $\Psi_0$ and $\Psi_4$ approach zero sufficiently rapidly, the scalars $\Psi_1$, $\Psi_2$ and $\Psi_3$ do not, unless $a$ is a constant. The condition on $a$ is actually weaker, which can be observed by considering the tortoise coordinate in the Kerr background. It is defined by the condition
\be
\dfrac{dr_*}{dr}=\dfrac{r^2+a^2}{\Delta}
\ee
and is given by
\begin{align}
r_*&=r+2 \sqrt{-a^2-M^2} \tan ^{-1}\left(\frac{M-r}{\sqrt{-a^2-M^2}}\right)\\
&\qquad +M \log \left(r^2-2 M r-a^2\right)\ .\nonumber
\end{align}
In the asymptotic region ($v\to\infty$, $r\to\infty$), this behaves as
\be\label{tort}
v(u,r)=u+2r+4M \log r+\mathcal{O}(r^{-1})\ ,
\ee
and this behaviour even in the dynamical gKV spacetimes since the asymptotic behaviour \eqref{tort} is preserved unless $r_0$ scales exponentially with retarded null time $v$. Thus
\begin{align}
  \pad_r a\big(v(u,r)\big)\sim \half a'\big(v(u,r)\big)\ , \\
  \pad^2_r a\big(v(u,r)\big)\sim \half a''\big(v(u,r)\big)\ .
\end{align}
As a result a gKV spacetime is future asymptotically flat if for large $v$ the angular momentum to mass ratio approaches a constant value at least as fast as $|a'|\propto 1/v$.

\subsection{Geodesics of Kerr--Vaidya spacetimes} \label{appd}
Here we  summarise some useful facts about timelike geodesics in the ingoing KV metrics. The Lagrangian  $L$  of a massive test particle is
\begin{align}
  {L}=&\frac{1}{2}\left[\bigg(1-\frac{2 M r}{\rho^2}\bigg)\dot v^2-2 \dot v \dot r- \frac{4 a M r \sin^2\theta}{\rho^2}  {\dot{v}\dot{\psi}} \right. \nonumber \\
&\left. -2 a \sin^2\theta  {\dot{r} \dot{\psi}}  \!  +\rho^2  {\dot{\theta}^2} + \frac{\Sigma^2}{\rho^2}\sin^2\theta  {\dot{\psi}^2} \right]\ .
    \label{kvL}
\end{align}
The Lagrange equations of motion is
\begin{equation}
    \frac{d}{d\tau}\left(\frac{\partial L}{\partial \dot x^\mu}\right)- \frac{\partial L}{\partial x^\mu}= 0\ ,
\end{equation}
which for $r$ component becomes
\begin{equation}
   -\frac{1}{2} \frac{d}{d\tau}(\dot v- a\sin^2\theta \dot\psi)- \frac{\partial L}{\partial r}= 0\ .
\end{equation}
Substituting the ZAMO condition~\eqref{zamo49} and simplifying further gives
\be
    -\frac{1}{2} \frac{d}{d\tau}\left(\frac{\rho^2}{\Sigma^2}\big((r^2+a^2) \dot v- a^2 \sin^2\theta \dot r\big) \right)- \frac{\partial L}{\partial r}= 0\ .\label{eqr}
\ee
Similarly, $v$ component of the Lagrangian equations of motion becomes
\begin{equation}
     -\frac{1}{2}  \frac{d}{d\tau} \left(\dot r-\left(1-\frac{2 M r}{\rho^2}\right)\dot v - \frac{2 a M r \sin^2\theta}{\rho^2} \dot\psi\right)- \frac{\partial L}{\partial v}= 0\ .
\end{equation}
Again substituting ZAMO condition~\eqref{zamo63} and simplifying further gives
\be
     \frac{1}{2} \frac{d}{d\tau}\left(\frac{\rho^2}{\Sigma^2} \big(\Delta\dot v- (r^2+a^2) \dot r\big) \right)  - \frac{\partial L}{\partial v}= 0\ . \label{eqv}
\ee
Multiplying Eq.~\eqref{eqr} by $\Delta$ and Eq.~\eqref{eqv} by $r^2+a^2$ and adding the equations eliminates the $\ddot v$ terms, allowing one to obtain the expression for $\ddot r$ in terms of first derivatives only
\begin{widetext}
    \begin{multline}
        \left(\dot v\Delta \frac{d}{d\tau}(r^2+a^2)- (r^2+a^2)\dot v \frac{d\Delta}{d\tau}- \Delta \dot r \frac{d}{d\tau} (a^2 \sin^2\theta)+ (r^2+a^2) \dot r \frac{d}{d\tau}(r^2+a^2)+ \Sigma^2 \ddot r\right) \frac{\rho^2}{\Sigma^4} \\
         + \dot r \frac{d}{d\tau} \left(\frac{\rho^2}{\Sigma^2}\right)- \Delta \frac{\partial L}{\partial r}- (r^2+a^2) \frac{\partial L}{\partial v}=0\ . \label{EOMv}
    \end{multline}
\end{widetext}
 Analysis of the leading order terms near the horizon shows that the acceleration is negative and decreasing as the particle goes outward from the expanding black hole:
\begin{equation}
    \ddot r \approx -\frac{4 r \left(a^2+r^2\right) M'(v)}{\Delta^2} \dot r^2+ {\cal O}(1/\Delta)\ .
\end{equation}
Following the same procedure for the evaporating white hole shows that the acceleration is negative and increasing as the particle approaches the evaporating white hole:
\begin{equation}
    \ddot r \approx \frac{4 r \left(a^2+r^2\right) M'(u)}{\Delta^2} \dot r^2+ {\cal O}(1/\Delta)\ . \label{d8}
\end{equation}


\begin{thebibliography}{}

\bibitem{CP:19} V.\ Cardoso and P.\ Pani, Testing the nature of dark compact objects: a status report.
{\href{https://doi.org/10.1007/s41114-019-0020-4}{ {Living Rev.\ Relativ.} \textbf{22}, 4  (2019)}}.

\bibitem{BCNS:19} L.\ Barack, V.\ Cardoso, S.\ Nissanke, and T.\ P.\ Sotiriou, Black holes, gravitational waves and fundamental physics: a roadmap. {\href{https://doi.org/10.1088/1361-6382/ab0587}{Class. Quant. Grav. \textbf{36}, 143001 (2019)}}.

\bibitem{F:14} V.~P.~Frolov, Information loss problem and a 'black hole` model with a closed apparent horizon.
	\href{https://doi.org/10.1007/JHEP05(2014)049}{JHEP \textbf{05}, 049 (2014).} [arXiv:1402.5446 [hep-th]].


\bibitem{MMT:22} R.\ B.\ Mann, S.\ Murk, and D.\ R.\ Terno, Black holes and their horizons in semiclassical and modified theories of gravity. {\href{https://doi.org/10.1142/S0218271822300154}{Int. J. Mod. Phys. D \textbf{31}, 2230015 (2022).}}


\bibitem{HE:73} S. W. Hawking and G. F. R. Ellis,
{\href{https://doi.org/10.1017/CBO9780511524646}{\textit{The Large Scale Structure of  Space-Time} (Cambridge University Press, Cambridge, England, 1973).}}

\bibitem{W:84} R. \ M. \ Wald,
{\href{https://doi.org//10.7208/chicago/9780226870373.001.0001}{\textit{General Relativity} (The University of Chicago Press, Chicago, 1984).}}


\bibitem{C:92} S. Chandrasekhar,
{\href{https://doi.org/10.1007/978-94-009-6469-3_2}{\textit{The Mathematical Theory of Black}
		Holes (Oxford University Press, Oxford, England, 1992).}}

\bibitem{O:95} B. O'Neil, \textit{The Geometry of Kerr Black Holes}, (Peters, Wellesley, MA, 1995).

\bibitem{FN:98}  V. P. Frolov and I. D. Novikov,
{\href{https://doi.org/10.1007/978-94-011-5139-9}{\textit{Black Holes: Basic Concepts and New Developments} (Kluwer, Dordrecht, 1998).}}

\bibitem{SKMHH:03} H. Stephani, D. Kramer. M. MacCallum, C. Hoenselaers, and E. Herlt,
{\href{https://doi.org/10.1017/CBO9780511535185}{\textit{Exact Solutions to Einstein's Field Equations} (Cambridge University Press, Cambridge, England, 2003).}}




\bibitem{vF:15}  V. Faraoni,
{\href{https://doi.org/10.1007/978-3-319-19240-6}{\textit{Cosmological and Black Hole Apparent Horizons}, (Springer, Heidelberg, 2015).}}


\bibitem{M:23}
S.~Murk, 
\href{https://doi.org/10.1142/S0218271823420129}{{Int. J. Mod. Phys. D \textbf{32}, in press (2023).}}



\bibitem{AB:05} A.\ Ashtekar and M.\ Bojowald,
		{\href{https://doi.org/10.1088/0264-9381/22/16/014}{{Class.\ Quantum Gravity} \textbf{22}, 3349 (2005)}}.	

\bibitem{A:20} A.\ Ashtekar,
		{\href{https://doi.org/10.3390/universe6020021}{ {Universe} \textbf{6}, 21 (2020)}}.	

\bibitem{F:05} L.\ H. \ Ford, Spacetime in Semiclassical Gravity,
\href{https://arxiv.org/abs/gr-qc/0504096}{arXiv:gr-qc/0504096 (2005).}

\bibitem{K:12}  K. Kiefer,  \textit{Quantum Gravity} (Oxford University Press, Oxford, 2012, 3rd edition).


\bibitem{DSST:24} P. K.  Dahal, F. Simovic,  I. Soranidis, I., and D. R.  Terno, 
\href{https://doi.org/10.1103/PhysRevD.110, 044032}{Phys. Rev. D \textbf{110}, 044032 (2024).}




\bibitem{MMT:22D} R.\ B.\ Mann, S.\ Murk, and D.\ R.\ Terno, Paradoxes before the paradox: setting up the information loss problem.
\href{https://doi.org/10.1103/PhysRevD.105.124032}{Phys. Rev. D \textbf{105}, 124032 (2022).}


\bibitem{DMT:22}
P.~K.~Dahal, S.~Murk and D.~R.~Terno, Semiclassical black holes and horizon singularities.
{\href{https://doi.org/10.1116/5.0073598}{AVS Quantum Sci. \textbf{4},  015606 (2022).} }

\bibitem{V:16} R. V. Vasudevan et al., A selection effect boosting the contribution from rapidly spinning black holes to the cosmic X-ray background.
{\href{https://doi.org/10.1093/mnras/stw363}{Mon. Not. R. Astron. Soc. \textbf{458}, 2012 (2016).}}

\bibitem{J:18} J. Jiang, \textit{et al.}, The 1.5 Ms observing campaign on IRAS 13224-3809-I. X-ray spectral analysis. {\href{https://doi.org/10.1093/mnras/sty836}{Mon. Not. R. Astron. Soc. \textbf{477}, 3711 (2018).}}

\bibitem{R:19} J. Roulet, M. Zaldarriaga, Constraints on binary black hole populations from LIGO–Virgo detections.
{\href{https://doi.org/10.1093/mnras/stz226}{Mon. Not. R. Astron. Soc. \textbf{484}, 4216 (2019).}}

\bibitem{AGLT:24}   C. Adamcewicz, S. Galaudage, P. D. Lasky, and E. Thrane,
{\href{https://doi.org/10.3847/2041-8213/ad2df2 }{Astr. J. Lett. \textbf{964}, L6 (2024).}}


\bibitem{MT:70} M. Murenbeeld and J. R. Trollope,  {\href{https://journals.aps.org/prd/abstract/10.1103/PhysRevD.1.3220} {Phys. Rev. D \textbf{1}, 3220   (1970)}}.
\bibitem{CK:77} M. Carmeli and M. Kaye, {\href{https://doi.org/10.1016/0003-4916(77)90263-9}{Ann. Phys. \textbf{103}, 97 (1977).}}
\bibitem{CGK:90} T. Christoulakis, T. Grammenos and C. Kolassis, 
 {\href{https://doi.org/10.1016/0375-9601(90)90892-R}{Phys. Lett. \textbf{149A}, 354 (1990).}}

\bibitem{B:62} H. Bondi, M. G. J. Van der Burg and A. W. K. Metzner,
{\href{https://doi.org/10.1098/rspa.1962.0161}{Proc. Roy. Soc. A \textbf{269}, 1336 (1962).}}

\bibitem{X:99} D.-Y. Xu,
    {\href{https://iopscience.iop.org/article/10.1088/0264-9381/16/2/002} {Class. Quant. Grav. \textbf{16}, 343  (1999).}}
\bibitem{ST:15} J. M. M. Senovilla and R. Torres,
 {\href{https://iopscience.iop.org/article/10.1088/0264-9381/32/18/189501}{Class. Quant. Grav.  \textbf{32}, 189501 (2015).}}
\bibitem{DT:20} P. K. Dahal and D. R.  Terno, 
{\href{https://doi.org/10.1103/PhysRevD.102.124032}{Phys. Rev. D \textbf{102}, 124032 (2020).}}



\bibitem{P:10} T. Padmanabhan,
{\href{https://stacks.iop.org/RoPP/73/046901}{Rep. Prog. Phys. \textbf{73}, 046901 (2010).}}
\bibitem{P:15} T. Padmanabhan, \textit{Gravity and the Spacetime: An Emergent Perspective}, in A. Ashtekhar and V. Petkov,
{\href{https://doi.org/10.1007/978-3-642-41992-8}{\textit{Springer Handbook of Spacetime} (Springer Berlin, Heidelberg, 2014).}}

\bibitem{DS:23} P. K. Dahal and F. Simovic, The Hawking temperature of dynamical black holes via Rindler transformations.
{\href{https://arxiv.org/abs/2304.11833}{[arXiv:2304.11833 [gr-qc]]}}

\bibitem{VP:73} P. C. Vaidya and L. K. Patel,
\href{https://doi.org/10.1103/PhysRevD.7.3590}{Phys. Rev. D \textbf{7}, 3590 (1973)}

\bibitem{york1983}
J. W. York, Jr., Dynamical Origin of Black Hole Radiance.  {\href{https://doi.org/10.1103/PhysRevD.28.2929}{Phys. Rev. D \textbf{28}, 2929 (1983).}}




\bibitem{P:76} D.\ N.\ Page, Particle emission rates from a black hole. II. Massless particles from a rotating hole
{\href{https://doi.org/10.1103/PhysRevD.14.3260}{ {Phys.\ Rev.\ D} \textbf{14}, 3260 (1976)}}.



\bibitem{DKS:16}    R. Dong, W. H. Kinney and D. Stojkovic,  
{\href{https://iopscience.iop.org/article/10.1088/1475-7516/2016/10/034} {J. Cosmol. Astropart. Phys. \textbf{10} 034 (2016)}}.

\bibitem{AAS:20}    A. Arbey, J.  Auffinger, and J. Silk, 
 {\href{https://academic.oup.com/mnras/article/494/1/1257/5810680}{Mon. Not. R. Astron. Soc. \textbf{494}, 1257 (2020)}}.

\bibitem{PDG:22} Particle Data Group,
{\href{https://doi.org/10.1093/ptep/ptac097}{Prog. Theor. Exp. Phys. \textbf{2022},  8 (2022)}}

\bibitem{DTAS:18} N. Dadhich, A. Tursunov, B. Ahmedov and  Z. Stuchlík, 
{\href{https://doi.org/10.1093/mnrasl/sly073}{Mon. Not. R. Astron. Soc. Lett. \textbf{478}, L89 (2018).}}

\bibitem{B:60} H. Bondi,
\href{https://doi.org/10.1038/186535a0}{Nature \textbf{186}, 535 (1960).}

\bibitem{S:62} R. Sachs,
\href{https://doi.org/10.1098/rspa.1962.0206}{Proc. R. Soc. London Ser. A \textbf{270}, 103  (1962).}

\bibitem{AF:13} M. A. Abramowicz and P. C. Fragile
{\href{https://doi.org/10.12942/lrr-2013-1}{Living Rev. Relativ. \textbf{16}, 1 (2013).}}

\bibitem{FEFHM:17} V. Faraoni,  G. F. R. Ellis, J. T. Firouzjaee, A. Helou, and I. Musco, Foliation dependence of black hole apparent horizons in spherical symmetry. {\href{https://doi.org/10.1103/PhysRevD.95.024008}{Phys. Rev. D \textbf{95}, 024008 (2017).}}

\bibitem{T:19} D.\ R.\ Terno, Self-consistent description of a spherically-symmetric gravitational collapse.
{\href{https://doi.org/10.1103/PhysRevD.100.124025}{Phys. Rev. D \textbf{100}, 124025 (2019).}}

\bibitem{BL:08} C. Barcelo, S. Liberati, S. Sonego and M. Visser, Fate of gravitational collapse in semiclassical gravity.
{\href{https://doi.org/10.1103/PhysRevD.77.044032}{Phys. Rev. D \textbf{77}, 044032 (2008).}}

\bibitem{MS:23}  S. Murk and I. Soranidis, Regular black holes and the first law of black hole mechanics.
{\href{https://doi.org/10.1103/PhysRevD.108.044002}{ {Phys.\ Rev.\ D} \textbf{108}, 044002 (2023).}}

\bibitem{sm:23} S.~Murk and I.~Soranidis, Kinematic and energy properties of dynamical regular black holes. \href{https://doi.org/10.48550/arXiv.2309.06002}{[arXiv:2309.06002 [gr-qc]].}

\bibitem{BMMT:19} V. Baccetti, R. B. Mann, S. Murk, and D. R. Terno, Energy-momentum tensor and metric near the Schwarzschild sphere. {\href{https://link.aps.org/doi/10.1103/PhysRevD.99.124014}{Phys. Rev. D \textbf{99}, 124014 (2019).}}

\bibitem{HWE:82} W. A. Hiscock, L. G. Williams, and D. M. Eardley, Creation of particles by shell-focusing singularities. {\href{https://doi.org/10.1103/PhysRevD.26.751}{Phys. Rev. D \textbf{26}, 751 (1982).}}

\bibitem{DST:22} P. K. Dahal, I. Soranidis, and D. R. Terno, Matter and forces near physical black holes. {\href{https://journals.aps.org/prd/abstract/10.1103/PhysRevD.106.124048}{Phys.\ Rev.\ D \textbf{106}, 124048 (2022)}}.

\bibitem{M:15} R.\ B.\ Mann,
{\href{https://doi.org/10.1007/978-3-319-14496-2}{\textit{Black Holes: Thermodynamics, Information, and Firewalls} (Springer, New York, 2015)}}.

\bibitem{WR:80} R. M.  Wald and S. Ramaswamy, Particle production by white holes.
{\href{https://doi.org/10.1103/PhysRevD.21.2736}{Phys. Rev. D \textbf{21}, 2736–2741 (1980).}}


\bibitem{FT:08} F.\ Fayos and R.\ Torres, A class of interiors for Vaidya's radiating metric: singularity-free gravitational collapse,''
{\href{https://doi.org/10.1088/0264-9381/25/17/175009}{Class. Quant. Grav. \textbf{25}, 175009 (2008).}}




\bibitem{BMT:18} V.\ Baccetti, R.\ B.\ Mann, and D.\ R.\ Terno, Role of evaporation in gravitational collapse. {\href{https://iopscience.iop.org/article/10.1088/1361-6382/aad70e/meta}{Class. Quant. Grav. \textbf{35}, 185005 (2018)}}.

\bibitem{heje:79} C. Gonz\'{a}lez, L. Herrera, and    J. Jim\'{e}nez, Radiating Kerr–Newman metric. {\href{https://aip.scitation.org/doi/pdf/10.1063/1.524156}{J. Math. Phys. \textbf{20}, 837 (1979).}}


\bibitem{KS:20}  E.-A.\ Kontou and K.\ Sanders, Energy conditions in general relativity and quantum field theory. {\href{https://doi.org/10.1088/1361-6382/ab8fcf}{Class. Quant. Grav. \textbf{37}, 193001 (2020)}}.













\bibitem{sS:11} S. N. Solodukhin, Entanglement Entropy of Black Holes.
{\href{https://doi.org/10.12942/lrr-2011-8}{Living Rev. Relativ. \textbf{14}, 8 (2011).}}

\bibitem{B:15} D. Buchholz and R. Verch, Macroscopic aspects of the Unruh effect. {\href{https://doi.org/10.1088/0264-9381/32/24/245004}{Class. Quant. Grav. \textbf{32}, 245004 (2015).}}

\bibitem{G:11} U. H. Gerlach, Absolute nature of the thermal ambience of accelerated observers.
{\href{https://doi.org/10.1103/PhysRevD.27.2310}{Phys. Rev. D \textbf{27}, 2310  (1983).}}


\bibitem{C:95} G. Cognola, L. Vanzo and S. Zerbini, One-loop quantum corrections to the entropy for a four-dimensional eternal black hole.
{\href{https://doi.org/10.1088/0264-9381/12/8/010}{Class. Quantum Grav. \textbf{12}, 1927  (1995).}}


\bibitem{B:13} E. Bianchi and A. Satz, Mechanical laws of the Rindler horizon.
{\href{https://doi.org/10.1103/PhysRevD.87.124031}{Phys. Rev. D \textbf{87}, 124031 (2013).}}

\bibitem{H:10} H. Chung, Asymptotic symmetries of Rindler space at the horizon and null infinity.
{\href{https://doi.org/10.1103/PhysRevD.82.044019}{Phys. Rev. D \textbf{82}, 044019 (2010).}}





\bibitem{BMPS:95} R.\ Brout, S.\ Massar, R.\ Parentani, and P.\ Spindel, A primer for black hole quantum physics. {\href{https://doi.org/10.1016/0370-1573(95)00008-5}{Phys. Rep. \textbf{260}, 329 (1995)}}.

\bibitem{Dorau:2024} P.~Dorau and R.~Verch,
\href{https://doi.org/10.1088/1361-6382/ad51c3}{Class. Quant. Grav. \textbf{41}, 145008 (2024).}

\bibitem{F:04} J. \ Frauendiener, Conformal Infinity,
{\href{https://doi.org/10.12942/lrr-2004-1}{Living Rev. Relativ. \textbf{7}, 1 (2004).}}

\bibitem{ANK:12} T. M. Adamo, E. T. Newman, C. Kozameh, Null Geodesic Congruences, Asymptotically-Flat Spacetimes
and Their Physical Interpretation,
{\href{https://doi.org/10.12942/lrr-2012-1}{Living Rev. Relativ. \textbf{15}, 1 (2012).}}


\end{thebibliography}
\end{document}